\documentstyle[aps,multicol,epsf]{revtex}
\draft
\def\beginwide{
        \end{multicols} \vspace*{-0.5cm} \noindent
        \rule{3.5in}{.1mm}\rule{.1mm}{5mm} \widetext \medskip }
\def\beginwidetop{
        \end{multicols} \vspace*{-0.5cm} \noindent
        \widetext \medskip }
\def\endwide{
        \hspace*{3.35in}~\rule[-5mm]{.1mm}{5mm}\rule{3.5in}{.1mm}
        \begin{multicols}{2} \vspace*{-1.0cm} \noindent }
\def\endwidebottom{
        \begin{multicols}{2} \vspace*{-1.0cm} \noindent }

\newcommand{\beq}{\begin{equation}}
\newcommand{\eeq}{\end{equation}}
\newcommand{\bdis}{\begin{displaymath}}
\newcommand{\edis}{\end{displaymath}}
\newcommand{\bea}{\begin{eqnarray}}
\newcommand{\eea}{\end{eqnarray}}
\newcommand{\barr}{\begin{array}}
\newcommand{\earr}{\end{array}}
\begin{document}

\title{How self-organized criticality works: A unified mean-field 
        picture}

\author{Alessandro Vespignani}

\address{International Centre for Theoretical Physics (ICTP)\\ 
P.O. Box 586, 34100 Trieste, Italy}

\author{Stefano Zapperi}
\address{Center for Polymer Studies and Department of Physics\\
Boston University, Boston, MA 02215}

\date{\today}

\maketitle
\begin{abstract}

We present a unified mean-field theory, based on the
single site approximation to the master-equation, for stochastic
self-organized critical models. In particular, we analyze in 
detail the properties of sandpile and forest-fire (FF) models.
In analogy with other non-equilibrium critical phenomena,
we identify the order parameter with the density of ``active'' sites
and the control parameters with the driving rates. 
Depending on the values of the control parameters, the system
is shown to reach a subcritical (absorbing)
or super-critical (active) stationary state. 
Criticality is analyzed in terms of the singularities 
of the  zero-field susceptibility. 
In the limit of vanishing control parameters, 
the stationary state displays scaling characteristic 
of self-organized criticality (SOC).
We show that this limit corresponds to the breakdown of space-time 
locality in the dynamical rules of the models. 
We define a complete set of critical exponents, describing
the scaling of order parameter, response functions,
susceptibility and correlation length
in the subcritical and supercritical states. 
In the subcritical state, the response of the system to small
perturbations takes place in avalanches. We analyze their scaling
behavior in relation with branching processes.
In sandpile models because of conservation laws,
a critical exponents subset displays
mean-field values ($\nu=1/2$ and $\gamma = 1$) in any dimensions.
We treat bulk and boundary dissipation and 
introduce a new critical exponent
relating dissipation and finite size effects. We present
numerical simulations that confirm our results.
In the case of the forest-fire model, our approach can
distinguish between different regimes (SOC-FF and deterministic FF)  
studied in the literature and determine the full spectrum
of critical exponents.

\end{abstract}


%
%
%
%

\begin{multicols}{2}

\section{INTRODUCTION}

After ten years of research and countless papers the precise
significance of self-organized criticality (SOC) \cite{btw}
is still controversial. Originally, SOC was 
presented as a general theory to understand 
fractals and $1/f$ noise as the natural outcome of 
the dynamical evolution of systems with many coupled 
degrees of freedom. Irreversible dynamics would
generate a self-organization of the system
into a critical state, without the fine tuning of external
parameters. The SOC idea was illustrated by 
computer models in which a slow external driving
leads to a stationary state with
avalanches of widely distributed amplitude \cite{btw}. 
This proposal stimulated a cascade of research activity
in experiments, 
theory and simulations. While the explanation
presented in Ref.~\cite{btw} about the origin of scaling in
nature appears now too simplistic, SOC gave a formidable input to  
the study of slowly driven systems and avalanche phenomena.

Avalanche behavior was experimentally observed in a variety 
of phenomena ranging from magnetic systems (the Barkhausen effect)
\cite{bark} and flux lines in high-$T_c$ superconductors \cite{flux},
fluid flow through porous media \cite{porous},
microfracturing processes \cite{ae},
earthquakes \cite{gr}, and lung inflation \cite{lung}.
In addition, SOC ideas stimulated a great interest in granular
matter \cite{sand}, although it was soon realized that the concept
was hardly applicable there, apart from the academic example
of a ricepile \cite{rice}. All the mentioned experiments share with SOC
models the slow external driving and the avalanche response,
but it is unclear whether self-organization as described in
Ref.~\cite{btw} plays any role in there. To answer this question it would
be necessary to better understand what determines the appearance
of scaling in SOC models and driven systems in general.

The idea of a critical point without fine tuning of external
parameters is very appealing because it opposes the
standard picture of equilibrium critical phenomena.
The concept of ``spontaneous'' criticality, as it has been
discussed in the literature, presents, however, several
ambiguities. It has been pointed out by several authors
that the driving rate is a parameter that has to be fine
tuned to zero in order to observe criticality \cite{grin,sor1,ddrg,vzprl}. 
This fact poses no problems
to computer simulations, where an infinite timescale separation
can easily be enforced, but it is crucial in experiments where the
driving rate is always non-zero. The second ambiguity is mostly a 
language problem: calling ``self-organization'' the evolution
towards a stationary state can be misleading. 
Any non-equilibrium system poised at its ``fine tuned'' 
critical point, when started from a generic configuration, evolves
towards the critical stationary state, thus building
up correlations and scaling. We would not describe this process
as self-organization.
These ambiguities in the definition of SOC have hindered
the formulation of precise relations with other non-equilibrium
critical phenomena \cite{zia}.

In the past years, several attempts have been made
to find a general mechanism to describe SOC models.
Sornette proposed that SOC was due to a 
non-linear feedback between order parameter
and control parameter, leading to the self-tuning
of the latter to the critical point \cite{sor2}. 
Later it was pointed out that this
mechanism could lead also to a first-order transition
rather than a critical point \cite{gil}. In a recent paper,
it was claimed instead that SOC corresponds to the
tuning to zero of the {\em order parameter} of an
ordinary critical phenomenon \cite{sor1}. Our analysis shows
that the situation is simpler: criticality
arises from the fine tuning to zero of one or more
{\em control parameters} (driving rate, dissipation)
and there is no coupling between control and order parameters \cite{vzprl}.
The incorrect identification of control
and order parameters is at the basis of many 
misconceptions about SOC phenomena, as we will discuss 
in the following.

Many  theoretical methods have been used in the analysis of SOC 
models. Few rigorous milestones can  be found in the activity of Dhar
and coauthors \cite{dhar,dhar2} and in the Dubna group works\cite{iv}.
Flory \cite{zhang} and Langevin-type approaches \cite{hwa,grin2,diaz} 
have been used on a phenomenological basis. More recently, real space 
renormalization group provided good estimates of the avalanche 
exponents \cite{ddrg,pvz}. Despite their richness, however, 
all these approaches are focused on the critical avalanche behavior and 
the external driving does not play any role; i.e. the system is 
studied in the infinite timescale separation regime. 
Furthermore, many of these attempts are conceived ad hoc for particular 
models and do not provide a general conceptual 
framework for SOC phenomena.

The first step towards a comprehensive 
theoretical understanding of SOC  
is provided by mean-field (MF) theory which gives 
insight into the fundamental 
physical mechanisms of the problem and a reference language.
It provides a feasible treatment to 
nonequilibrium and complex  problems (often the only one), 
and can be used as a starting point for 
more sophisticated calculations.
Whereas many 
numerical and analytical approaches get harder as 
the dimensionality increases, MF theory improves and
despite the crude approximations it
usually gives correct qualitative predictions for the 
phase diagrams of high-dimensional systems.  
Finally, MF theory highlights the
importance of symmetries and conservation laws.

A vast activity concerning MF theory of SOC 
models can be found in the literature. Exponents
describing avalanche distributions and propagation have been computed
in several ways: solving infinite-range \cite{jl}, Bethe lattice \cite{dhar2}
and random neighbor \cite{fbs,mfff,co}  
models and by mapping the dynamics into a branching process
\cite{alstr,gp,sobp,sobp2,amos}. Self-consistent MF approximations 
for the sand height distribution have also been used \cite{zhang,katori}. 
Other MF approaches use analogies with equilibrium
critical phenomena \cite{bakmf,stellamf}, 
leading sometimes to incorrect predictions
as we will discuss in the following. In summary, the MF
approach to SOC systems is composed of a number of studies
of specific models but a comprehensive understanding
of the phenomenon is missing. 

Here, we present a unified MF description of SOC models 
using the formalism developed for non-equilibrium critical
phenomena with steady states. We use a single site approximation
to the master-equation and we enforce conservation
laws by effective parameters and constraints. We concentrate
on models driven by stochastic noise, 
such as the sandpile \cite{btw} and the forest-fire (FF) \cite{dro}. 
In order to write a master equation, we consider 
finite values of the driving rates, since only in
this case the dynamical rules are local in space and time.
 Our analysis shows that
criticality in these models corresponds to the limit in which
the dynamical rules become non-local. 
Non-locality is implicitly enforced in computer simulations,
where the evolution of a single site 
depends on the state of the entire system. 
This fact is particularly evident in extremal models
in quenched disorder, where the dynamics 
proceeds by a global minimum search. Also in that case,  
to write a local master-equation one has to introduce a non-zero
driving rate, but the driving mechanism differs from the one we discuss
here and will be described in a forthcoming paper \cite{loreto}.

The present approach allows us to identify control 
and order parameters of SOC models and to clarify
the relations with other non-equilibrium critical 
phenomena \cite{vzprl}. In particular, we show that SOC models
have close similarities with non-equilibrium cellular
automata with many absorbing states \cite{torre,dick,jmendes}. 
The major difference
is that in SOC models, the control parameters have to be tuned
to zero to reach criticality. As we discussed before, this 
limit corresponds to the breakdown of locality in the 
dynamical rules of the model and hence to the onset of long
range correlation in the dynamical response. 
While apparently there is no mathematical difference between tuning
a parameter to zero or to a non-zero value, the physical
differences are quite important \cite{grin}. In the first case, 
changing the value of the control parameter by a given 
factor still keeps the system close to the critical point.
This is not the case in ordinary phase transitions, where, 
doubling the value of the temperature, the system looses
completely the critical properties. Moreover, in order 
for SOC model to be defined, the control parameter (i.e. the driving
rate) should always be non-zero and the critical point can only 
be reached through a limit process. 

The present MF theory can be applied to any stochastic
cellular automaton and therefore provides a unified description
for the ensemble of SOC models and other related non-equilibrium 
critical systems,
such as contact processes and cellular automata with absorbing states
\cite{torre,dick,jmendes}.
Moreover, it serves to emphasize 
the differences between different models and between different
regimes in the same model. Our analysis points also out 
the inconsistencies contained in earlier MF 
approaches\cite{bakmf,stellamf} which led to a misleading characterization 
of the model. We identify the sub-critical and
super-critical states of SOC models and discuss the different
ways in which criticality can be reached. 
We describe the avalanche behavior characteristic 
of these models in terms of response functions 
and we study the effect of perturbations on the stationary state.
We introduce a full set of critical exponents, describing
the response at the critical point and the scaling
close to the critical point in the sub-critical and super-critical 
states. In the case of sandpile models a subset of
exponents is found to have mean-field values in any dimension.
The reason for this behavior, which is confirmed by numerical
simulations, is ascribed to the presence of conservation
laws in the dynamics. 

The paper is organized as follows:
In Section II we introduce the models.
In Section III we review the dynamic mean-field approximation
to the master equation. Section IV contains the mean-field theory
for the sandpile model and discusses some issues 
related to conservation. In Section V we report the mean-field
analysis for the FF model and Section VI 
is devoted to a general discussion. A brief report of these results 
has appeared in Ref.~\cite{vzprl}.

\section{THE MODELS}

A rapid look at the SOC literature discourages every newcomer in the 
field. In less than ten years more than two thousand papers have been 
published, and 
comprehensive reviews are not yet appeared (a valuable effort in 
this direction can be found in Ref.~\cite{grin}). This is 
due to the lack of a general understanding  which would 
provide the framework to order the huge amount of informations 
about SOC. In particular, we were spectators of an hectic activity in 
numerical simulations, with the introduction of a multitude of different
models. A closer look at the literature reveals that the number 
of original models can be greatly reduced by noting 
that most of them are variations of prototype models. 
Using a  more ``Draconian'' approach, we can  
distinguish just two main families of SOC models.
The first is represented by the stochastic SOC models such as the sandpile or 
forest-fire model, in which the self-organization process is the output
of a stochastic dynamics. The second family groups together the 
so-called ``extremal'' or ``quenched'' models \cite{extrem}, 
which are defined by a 
deterministic dynamics in a random environment. Examples of the latter
family are the Invasion Percolation (IP) \cite{ip} 
and the Bak-Sneppen (BS) \cite{bs} models.
In this paper we discuss stochastic models, but work is in 
progress to extend the present analysis to systems driven by an extremal 
dynamics. 

\subsection{Sandpile models}

Sandpile models are cellular automata (CA) with an integer
(or continuous) variable $z_i$ (energy) defined in a $d-$dimensional
lattice. At each time step
an energy grain is added to a randomly chosen site, until
the energy of a site reaches a threshold $z_c$.
When this happens the site relaxes 
\beq
z_i\to z_i -z_c
\eeq 
and  the energy is transferred
to the nearest neighbors 
\beq
z_j\to z_j +y_j.
\eeq
The relaxation of a site can induce nearest neighbor sites to relax 
on their turn, i.e. they exceed the threshold because of the energy received.
New active sites can generate other relaxations and so on, eventually 
giving rise to an avalanche.
For conservative models the
transferred energy equals the energy
lost by the relaxing site ($\sum y_j=z_c$), at least on average.
Usually, the only form of dissipation occurs at
the boundary, from which energy can leave the system.
It is worth to remark that during the avalanches, the energy input stops
until the system is again in equilibrium and no active sites are presents.
This corresponds to an infinite timescale separation.
With these conditions the system reaches a stationary
state characterized by avalanches whose sizes $s$ 
are distributed as a power law \cite{btw,kada,grasma,manna1,lubeck}
\beq
P(s)\sim s^{-\tau}.
\eeq
The model originally introduced by Bak, Tang and Wiesenfeld (BTW)\cite{btw} 
is a discrete  automaton in which $z_c=2d$ and $y_j=1$.
An interesting variation  of the original sandpile is the 
three-states Manna model\cite{manna}. 
In this automaton the critical threshold is $z_c=2$ independently on the 
dimensionality $d$  and  if a relaxation (toppling) takes place, the energy
is distributed into two randomly chosen nearest neighbor sites. 
Other variations in which part of the energy is 
kept by the relaxing site can also be considered as well as directed 
models in which energy is transferred along 
a preferential direction\cite{kada}. 
Finally, sandpile models that include a relaxation 
dynamics where part of the energy is dissipated have 
been considered \cite{diss}. These models can be characterized by the fraction
of energy that disappears from the system during each relaxation process.
When a global dissipation is present (energy is lost on average),
the critical behavior is destroyed and a characteristic length is 
introduced. These numerical evidences suggest that 
conservation is necessary to obtain criticality.

As we discussed above, sandpile models are driven by 
adding a single energy grain on a 
randomly chosen site, when no active site is present. 
In this way, avalanches are instantaneous with respect to 
the driving timescale. This rule is very naturally implemented
in a computer algorithm, which can handle at the same time the two 
different timescales.
This, however, corresponds to a nonlocal interaction in which the site
dynamical evolution depends upon the whole system configuration. 
This non-local interaction is hard to describe, and 
in order to perform an analytical description we have to fix a 
reference timescale, as for example the single site relaxation step,
and measure the driving rate on that scale. For this reason,
we consider a generalized  sandpile model\cite{bakmf}, that  includes a 
non vanishing driving rate by  introducing the probability $h$ 
per unit time that a site will receive a grain of energy. Energy
is distributed homogeneously and the total energy flux is given by
$J_{in}=hL^d$. The parameter $h$ 
sets the driving timescale or equivalently the typical
waiting time between different avalanches as $\tau_d \sim 1/h$.
In the limit $h\to 0$, we recover the slow driving limit; i.e. during an 
avalanche the system does not receive 
energy. 
This formulation of the dynamics 
has the advantage to be local in space and time. The state 
of a single site depends only on the state of the site itself and its 
nearest-neighbor 
sites at the previous time step, through a transition probability
that is given by the reaction and driving rates. 
It will be also convenient in the ensuing analysis to group the possible
states of a site in three classes: active when $z \ge z_c$,
critical when $z=z_c-1$ and stable for $z < z_c-1$.

\subsection{Forest-Fire model} 

The first example of a stochastic 
SOC model without conservation is in the forest fire (FF) 
model \cite{dro}. The model has
been first introduced by Bak et al. \cite{bct} as an 
example of SOC, and has been 
then modified by Drossel and Schwabl \cite{drosch}.
The model is defined on a lattice in which
each site can be empty, occupied by a green tree 
or by a burning tree. Burning trees turns to ashes with
unitary rate and set fire to the neirest neighbor trees.
The model was first studied in the case of a small tree 
growth rate $p$ and in the absence of a spontaneous ignition of fires. 
In $d=2$ the system reaches a dynamical state in which fire fronts 
propagate with trivial scaling properties \cite{graka}. 
Only recently new large 
scale simulations have shown that for $d>2$ anomalous scaling 
laws occur \cite{bro}.
A more interesting situation appears when a very small rate for spontaneous 
fire ignition $f$ (lightning probability) is introduced in the automaton 
dynamics. The system shows scaling behavior with a diverging characteristic 
length in the limit $f/p\to 0$ and $p\to 0$ and the activity occurs in bursts 
of fire spreading (avalanches) whose distribution follows a power law behavior 
$P(s)\sim s^{-\tau}$.
In the FF model 
the driving rates are  explicitly defined by $f$ and $p$, and
the dynamical rules are thus local. However, in numerical simulations 
the two driving field are implicitly set to zero by the condition that 
trees growth and fire ignition occur only when the system does not show 
active sites. Only the ratio $\theta=f/p$ is quantitatively defined by the 
relative probability of tree growth with respect to  fire ignition events. 
Also in this case numerical simulations are done in the 
infinite timescale separation limit, which corresponds to the subcritical 
state of the system. 

It is interesting to note the similarity between SOC models and 
nonequilibrium lattice automata with multiple adsorbing states \cite{torre}. 
These models present a critical phase transition separating two regimes:
above the transition there is a finite density of active sites, 
while below the transition point this density is zero and 
the system freezes in one of the many stable configurations. 
In the following, using the formalism developed for this class
of models we will make this analogy more precise. 

\section{DYNAMIC MEAN FIELD APPROXIMATION}

In the previous section we generalized the definition of SOC automata to 
the fast-driving regime, thus removing the assumption of timescale separation 
commonly employed in simulations. This will turn out to be particularly 
convenient since the restored locality of the 
dynamical rules allows a simpler description of the models.

The most generic description of SOC models is through a $d$-dimensional
stochastic cellular automaton with $N=L^d$ sites, where $L$ is the 
lattice size. Each site $i$ on the lattice is characterized by an occupation
variable $\sigma_i$ which can assume $q$ different values: for instance
the possible energy levels $z_i$ or the  three different FF model states.
The complete set $\sigma=\{\sigma_i\}$ of lattice variables specifies a 
configuration of the system. The dynamical evolution of the system is 
determined by the transition probability $W(\sigma\mid\sigma')$ 
from the configuration $\sigma'$ to the configuration $\sigma$. 
At each time step the state of a given site depends only 
on the previous state of the site itself and the set 
of sites interacting with it. 
The most general transition probabilities in the homogeneous and symmetric 
case is 
\beq
W(\sigma\mid\sigma')=\prod_{i=1}^N w(\sigma_i\mid\sigma') 
\label{eq:W}
\eeq
where $w(\sigma_i\mid\sigma')$ is the one site transition 
probability that depends on driving and reaction rates. 
The single site transition probability should satisfy the 
normalization property
\beq
\sum_{\sigma_i}w(\sigma_i\mid\sigma')=1.
\eeq
Because of the intrinsic nonequilibrium behavior of these systems, 
we have to consider the time-dependent 
probability distribution $P(\sigma,t)$ to have 
a configuration $\sigma$ at time $t$. From this distribution 
we can compute the average value of any 
function of the state $A(\sigma)$  
\beq
\langle A(t) \rangle=\sum_{\sigma}A(\sigma)P(\sigma,t).
\eeq
The time evolution of the probability distribution is governed by the master 
equation (ME), which in continuous time reads as 
\beq
\frac{\partial}{\partial t}P(\sigma,t)=\sum_{\sigma'}W(\sigma\mid\sigma')
P(\sigma',t) - W(\sigma'\mid\sigma)P(\sigma,t).
\eeq
The specific form of $W$ determines the dynamics of the model
and the steady state distribution.
Typically, SOC systems show a stationary state in which all the single time 
averages are time independent. To this state corresponds a stationary 
probability distribution $P(\sigma)=P(\sigma,t\to\infty)$. For equilibrium 
systems the stationary distribution has the Gibbs form 
$P(\sigma)\sim\exp(-\beta H(\sigma))$, where $H(\sigma)$ is the Hamiltonian.
For SOC systems, like other nonequilibrium systems, there is not such a 
general criterion, but we have to solve explicitly
the ME in the stationary limit. 
In practice, this is a formidable task which is accomplished 
just in very few cases.
It is then necessary to use approximate 
methods in order to describe the collective behavior of these systems.
The simplest available method is the dynamic cluster variation 
approach, which involves a hierarchy of evolution equations for the 
probability distribution of configuration of cluster of $k$ sites: 
$P_k(\sigma_1,\cdots,\sigma_k)$. If the system is homogeneous, the 
distribution of cluster of $k$ sites will be position independent.
It is easy to recognize that $P_1$ represents 
the average density of sites in a certain state, 
while $P_{k>1}$ characterize the 
correlation properties of the systems. Unfortunately, the 
time evolution equations 
for each of the $P_k$ depends on the higher correlation 
functions: the dynamical 
equation for the average densities depends on the two 
point correlation functions,
the two points correlations on the three point 
correlations and so on. 
We have therefore an infinite chain of coupled equations. 
The dynamical mean-field approximation 
consists of neglecting correlations up to a certain order. 
In the {\em n-sites
approximation}  cluster probabilities are decoupled as a 
product of n-sites 
probabilities. This approximation has
proved to be quite instructive 
for a qualitative description of the 
critical behavior of nonequilibrium systems\cite{dick}.

Before proceeding to discuss in detail the single-site 
MF approximation for stochastic 
SOC models, we first discuss the basic symmetries of these systems, 
which will play a fundamental role in 
formulating a common description.
We can reduce the number of states each site in 
the system can 
assume, noting that we can always identify three main 
states: {\em stable} ($\sigma_i=s$), 
{\em critical} ($\sigma_i=c$) and 
{\em active} ($\sigma_i=a$). Stable sites are those that 
do not relax (become active)
if energy is added to them by external fields or interactions 
with active sites.
Critical sites become active with addition of energy. 
Active sites are those 
transferring energy; they are interacting with other sites 
(usually nearest neighbors).
Indeed, SOC refers always to systems in which the only 
state that  generates dynamical 
evolution is the active one; i.e. stable and critical 
sites can change their state only
because of external fields or by interacting with 
an active nearest neighbor. 
Therefore, SOC model correspond to three states 
CA on $d$-dimensional lattices. This description is only 
approximate, since a certain amount of 
information is lost in grouping together stable sites. 
For instance, in the BTW model\cite{btw} 
we have several energy level which pertain to a stable 
site, but we can take this fact into 
account introducing in the ME some effective 
parameters. The three states description is exact for 
the Manna model \cite{manna} and the FF model \cite{dro}.

In the simple MF single-site approximation, 
we denote by $\rho_a,\rho_c$ and $\rho_s$ 
the average densities of sites in the active, 
critical and stable states respectively. In the case 
of homogeneous systems,
these densities can be written as:
\beq
\rho_{\kappa}(t)=\sum_{\{\sigma\}}\delta(\sigma_j-\kappa)P(\sigma,t).
\eeq 
The dynamical equations for the average 
densities  are obtained from the ME by using Eq.~(\ref{eq:W}) 
\beginwide
\beq
\frac{\partial}{\partial t}\rho_{\kappa}(t)
=\sum_{\{\sigma'\}}\sum_{\{\sigma\}}\delta(\sigma_j-\kappa)(\prod_i 
w(\sigma_i\mid\sigma')P(\sigma',t) - 
\prod_i w(\sigma_i'\mid\sigma)P(\sigma,t)).
\label{me2}
\eeq
\endwide 
The above equation can be simplified by using the 
normalization condition for the 
transition probabilities
\beq
\sum_{\{\sigma'\}}\prod_i w(\sigma_i'\mid\sigma)=1,
\eeq
\beq
\sum_{\{\sigma\}}\delta(\sigma_j-\kappa)\prod_i w(\sigma_i\mid\sigma')=
w(\sigma_j=\kappa\mid\sigma').
\eeq
Equation (\ref{me2}) can be further simplified when the interactions
are only among a finite set of sites. In this case,
with $\sigma'=\{\sigma'_i,\sigma'_{i+e}\}$ we denote the site $i$
and the set of sites that can 
interact with it --- usually a finite number of sites
or more commonly just  the nearest-neighbors (n.n.). 
By restricting the sum and dropping the site index, 
because of the homogeneity, we 
finally get 
\beq
\frac{\partial}{\partial t}\rho_\kappa(t)
=\sum_{\{\sigma'\}}w(\kappa\mid\sigma')
P(\sigma',t) -
\rho_\kappa(t).
\eeq
It is worth to remark that in the above expression 
the set $\sigma'=\{\sigma'_i,\sigma'_{i+e}\}$ refers 
to the generic set of interacting sites which 
depends upon the particular dynamical rules and 
lattice geometry. In presence of a non-local
interaction, the set $\sigma'$ can correspond to the entire
system. This presents a very  
difficult problem that can be treated introducing a 
suitable regularization. 

In general, we have therefore that the 
evolution equations of the average 
densities are still coupled to the probability distribution 
of configurations
of a set of interacting  sites. 
In order to have a set of closed equation for the 
densities we truncate the 
evolution equations by using a dynamical MF single-site approximation.
In this  MF scheme we approximate the probability of 
each configuration $\sigma$ 
as the product measure of single-site probabilities
\beq
P(\sigma)=\prod_i P(\sigma_i)\equiv\prod_i\rho_{\sigma_i},
\eeq
thus neglecting all correlations in $P(\sigma)$. 
Introducing this approximation
in Eq.~(\ref{me2}), we obtain the 
MF reaction rate
equations which depends just on the single-site 
densities and can be symbolically 
written as  
\beq
\frac{\partial}{\partial t}\rho_\kappa=F_\kappa(\rho_a,\rho_c,\rho_s)~~~~~
~~~~~~~\kappa=a,c,s,
\label{ssmf}
\eeq
where $F_\kappa$ depends upon driving fields and  interactions 
parameters through the transition rates $w$. In addition, 
because the densities 
must preserve normalization, two 
of the above equations supplemented with the condition 
$\rho_a+\rho_c+\rho_s=1$, are enough to describe completely the system.

In practice the form of the 
rate equations depends upon the specific model. 
Nevertheless, we can write the general structure of the 
equations describing SOC models by simple considerations.
In general, $F_\kappa$ can be expanded as a 
series of the average densities:
\beq
F_\kappa= \sum_n f_\kappa^n\rho_n +\sum_{n,\ell}f_\kappa^{n,\ell}
\rho_n\rho_\ell + {\cal O}(\rho_n^3),
\eeq 
where the constant term is set to zero in order to get a stationary 
state. The first order terms are the transition rates generated by
the external driving fields or by spontaneous transitions. 
The second and higher 
order terms characterize  transitions due to the 
interaction between different sites. 
In SOC models, only the active state generates a 
non trivial dynamical evolution, 
while stable or critical sites can change 
their state only because of the external field or the presence of an active 
n.n. site. Since the critical point is identified by $\rho_a = 0$, 
in correspondence with a vanishing external field,
we can neglect second order terms in the density of active sites.
The solutions of the stationary equations
($\frac{\partial}{\partial t}\rho_\kappa=0$)
are function of the effective parameters $f_\kappa^n, f_\kappa^{n,\ell}$,
which depend on the details of the model.
It is expected that the critical behavior is not affected by the 
specific values of the parameters, while
universality classes will depend on constraints 
imposed on the equations because of symmetries and  conservation laws.

\section{MEAN-FIELD ANALYSIS OF SANDPILE MODELS}

We consider here the explicit application of the single site MF 
approximation to the class of sandpile models. A simpler derivation
based on symmetry considerations can be found in Ref.~\cite{vzprl}.
In the previous section we have shown that the 
MF dynamical equations reduce in this approximations 
to the the following expression
\beq 
\frac{\partial}{\partial t}\rho_\kappa(t)
=\sum_{\{\sigma'\}}w(\kappa\mid\sigma')
\prod_{i}\rho_{\sigma'_i}(t) -
\rho_\kappa(t).
\label{mfgen}
\eeq
where $\sigma'$ denotes the set formed by a single site and its  
set of interacting sites as specified by the dynamical rules. 
All the dynamical information of the system
is contained in the transition rates $w(\kappa\mid\sigma')$.
Unfortunately, the sandpile model is inherently non-local because of the 
implicit timescale separation. A site 
can receive energy only if the system is quescient. This implies that 
transition rates depend upon the whole set of lattice variables present in 
the system, giving rise to a strongly non-local dynamical rule.
In order to treat the model analytically we have to regularize this 
interaction by a suitable parametrization which allows to recover 
the non-locality in some particular limit. The simplest regularization 
has been discussed in Sec. II, introducing the external flow  of 
energy added to the system.  We describe this external flow by   
the probability per unit time $h$ for a site to receive 
a grain of energy. The transition rates are now local,  
depending only on the field $h$ and the state of the nearest neighbor 
sites ($\sigma_{n.n.}$) that determine the toppling dynamics.  
The total amount of energy 
added to the system at each time step will be  $J_{in}=h L^d$\cite{nota1}.
The non-locality of the dynamical rules is recovered in the limit $h\to 0$
(see Sec. VI), that corresponds to an infinite timescale separation.
The external field $h$ has been historically introduced in Ref.\cite{bakmf}.
Unfortunately, these 
early papers failed to address consistently the role played 
by driving and conservation, and led to several inconsistencies 
(see Sec. IV B). 
Other regularization can be introduced as well in the ME treatment. 
We limit our discussion to the present one for reasons of simplicity.
Nevertheless a more accurate characterization of the degree of non-locality 
actually present in the infinitely slowly driven sandpiles can be 
obtained via more refined regularization schemes \cite{vz}.

Since locality is restored ($\sigma'=\{\sigma'_i,\sigma'_{n.n.}\}$), 
we can derive the MF equations for the density of active sites
by considering the leading order in $h$ and $\rho_a$ in
Eq.~(\ref{mfgen}). The transition rates obey  
$w(a\mid a,\sigma_{n.n}) =0$, because an active site always transfers
its energy, thus  
becoming stable at the next time step independently
from its $n.n.$ sites. In this way, we are neglecting higher contribution 
due to the presence of multiple active  $n.n$ sites, which can transfer 
energy to the active site sustaining its activity. 
The only allowed transition to the active state are due to critical 
sites which receive energy from the external driving or from active $n.n.$
sites. In the absence of active $n.n$ sites, we have 
$w(a\mid c,\sigma_{n.n.}\neq a) =h$. We can then obtain
the contribution to the  dynamical MF equation 
\beq
\sum_{\{\sigma'_{n.n.}\}}w(a\mid c,\sigma'_{n.n.}\neq a)
\rho_c\prod_{i\in n.n.}\rho_{\sigma'_i}(t)=h\rho_c(1-\rho_a)^Z,
\label{eq:rhoa1}
\eeq
where $Z$ represent the lattice coordination number; i.e. the number of 
$n.n.$ sites. If the sites does not receive energy from
outside we have to consider the possibilities that one of the $n.n.$ sites
is active and transfers energy to it. This process corresponds to 
\beq
w(a\mid c,\sigma_{i}=a,\sigma_{j\neq i}\neq a) = (1-h)\frac{g}{Z}
(1-\tilde{p}),
\eeq
where $i,j \in n.n.$. The right term represent the probability that 
a critical site receive an energy grain only from an active $n.n.$.
This is equal to the ratio between the number of sites $g$ involved 
in the dynamical relaxation process and the total number 
of $n.n.$. For instance, $g=2d$ for the BTW model\cite{btw} or $g=2$
for the Manna model\cite{manna}.
In addition, we have to consider the probability 
$\tilde{p}$ that the active site does not transfer its energy because 
of intrinsic dissipation  or because it is a boundary site. 
The above transition rate is valid only for homogeneous processes
and therefore excludes directed models. 
The total contribution due to 
this process, considering the multiplicity of active $n.n.$ sites, is given
by
\bea 
\nonumber
\sum_{\{\sigma'_{n.n}\}}w(a\mid c,\sigma'_{i}=a,\sigma'_{j\neq i}\neq a)
\rho_c\rho_a\prod_{j\neq i\in n.n.}\rho_{\sigma'_j}(t)\\
=(g-\epsilon)
\rho_c\rho_a(1-h)(1-\rho_a)^{Z-1},
\label{eq:rhoa2}
\eea
where the
parameter $\epsilon$ conveniently identifies the average 
energy dissipated $g\tilde{p}$ in each elementary process.
It is worth to remark that $\epsilon$ is present also for 
fully conservative systems,
being an effective term due to the boundary dissipation:
it acts as an external tunable parameter in 
the case of bulk dissipation and accounts for size effects 
in finite systems.  

Neglecting higher orders in $h$ and $\rho_a$ from the Eqs.~(\ref{eq:rhoa1})
and (\ref{eq:rhoa2}),
we can finally write the MF dynamical equation for the densities of 
active sites  
\bea 
\nonumber
\frac{\partial}{\partial t}\rho_a(t)= -\rho_a(t) +h\rho_c(t)+\\
(g-\epsilon)\rho_c(t)\rho_a(t)+{\cal O}(h\rho_a,\rho_a^2).
\label{mf1}
\eea

Next, we derive the dynamical MF equation for the density
of stable sites, following the same strategy used above. 
Since, at lowest order, active sites become stable with unitary rate, 
we have that 
$w(s\mid a,\sigma'_{n.n.})=1+{\cal O}(h\rho_a,\rho_a^2)$, 
yielding a contribution to the MF equation which is 
\beq 
\sum_{\{\sigma'_{n.n}\}}w(s\mid a,\sigma'_{n.n.})
\rho_a\prod_{i\in n.n.}\rho_{\sigma'_i}(t)= 
\rho_a +{\cal O}(h\rho_a,\rho_a^2). 
\eeq
Since critical sites never become stable, 
we have also that $w(s\mid c,\sigma'_{n.n.})=0$. 

Energy conservation imposes that energy is stored 
in stable sites until they become critical. This implies a non unitary 
rate $w(s\mid s,\sigma'_{n.n.})$. The simplest way to derive this term  
make use of the normalization condition that yields
$w(s\mid s,\sigma'_{n.n.})=1-w(c\mid s,\sigma'_{n.n.})$.
In fact, these transition rates are nearly equivalent to those
from critical to active sites. The only difference is that only a fraction 
$u$ of stable sites receiving an energy quantum will contribute to the 
$s\to c$ process, i.e. the fraction of stable sites which are 
sub-critical. Therefore the reaction rates are related by the factor $u$
as $w(c\mid s,\sigma'_{n.n.})=u w(a\mid c,\sigma'_{n.n.})$. 
Recalling the derivation of Eq.~(\ref{mf1}),
it is straightforward to obtain 
\bea
\nonumber
\sum_{\{\sigma'_{n.n}\}}(1-w(c\mid s,\sigma'_{n.n}))
\rho_s\prod_{i\in n.n.}\rho_{\sigma'_i}(t)= \\
\rho_s -uh\rho_s
-u(g-\epsilon)\rho_s\rho_a +{\cal O}(h\rho_a,\rho_a^2). 
\eea
Adding all these contributions, we finally obtain the dynamical MF
equation 
\bea 
\nonumber
\frac{\partial}{\partial t}\rho_s(t)= \rho_a(t) -uh\rho_s(t)\\
+u(g-\epsilon)\rho_s(t)\rho_a(t)+{\cal O}(h\rho_a,\rho_a^2),
\label{eq:rhos}
\eea 
that together with Eq.(\ref{mf1}) and supplemented with the 
normalization condition fully describes the MF evolution of
sandpile automata. 

In deriving the MF equations we have made an approximation,
introducing the parameter $u$ to take into account
the presence of several energy levels instead of a single stable 
level. For three levels models $u=1$ and this description 
is exact, while for multilevel models like the BTW \cite{btw}
the parameter $u$ can be determined self-consistently 
in the stationary state using
the energy conservation \cite{vzprl}. Here we show that
we can also obtain $u$ by the full 
description of the MF equations. 
We consider a generic sandpile model
in which the energy threshold is $z_c$ and, after a relaxation event,
$g$ energy grains are transfered 
to randomly chosen neighbors. For instance, the $d$-dimensional BTW has 
$z_c=2d$ and $g=2d$, but we can think to arbitrary values 
for these parameters. 
We can describe in more details these systems by introducing the 
densities $\rho_n$, describing the probability that a site is in the level 
$n$. We then have that $\rho_c=\rho_{z_c-1}$ and 
$\rho_s=\sum_{n=0}^{z_c-2}\rho_n$. The dynamical evolution can be simplified 
noting that an active site with energy $z_c$ becomes stable 
and  its energy becomes $n=z_c-g$. One can show that
stable sites with energy levels 
lower than $n$, have a zero stationary density.
Without loss of generality,
we can therefore assume that the zero energy level 
is $n=z_c-g$. By rescaling the energy levels
 in this way, we get $\rho_c=\rho_{g-1}$ and 
$\rho_s=\sum_{n=0}^{g-2}\rho_n$. The intermediate level are described 
by the following set of MF equation 
\bea
\nonumber
\frac{\partial}{\partial t}\rho_n(t)= -h\rho_n(t)
-(g-\epsilon)\rho_n(t)\rho_a(t) +h\rho_{n-1}(t) \\
+(g-\epsilon)\rho_{n-1}(t)\rho_a(t)
+{\cal O}(h\rho_a,\rho_a^2),
\eea
where  $1\leq n\leq g-2$. 
In the stationary state we obtain 
$\rho_n=\rho_{n-1}\cdots=\rho_0$ and, noting that 
$u=\rho_{g-2}/(\sum_{n=0}^{g-2}\rho_n)$, we recover 
the result $u=1/(g-1)$ obtained in \cite{vzprl}. 
This result expresses the energy
conservation and fixes the stationary solution consistently 
with the energy balance. Noticeably, in the Manna model\cite{manna}
for which $g=2$, we obtain  $u=1$ 
as it must be for a three states model.  
As a last remark we point out that 
summing up the above set of  equations with the one for $\rho_0$ 
\bea
\nonumber
\frac{\partial}{\partial t}\rho_0(t)= -h\rho_0(t)+\\
(g-\epsilon)\rho_0(t)\rho_a(t) +\rho_a(t), 
\eea
we obtain Eq.~(\ref{eq:rhos}) as a function of the 
parameter $u$. 

When the system is far from the stationary state, 
the parameter $u$ will be in general time dependent. 
In the following we will always consider stationary properties 
or homogeneous perturbations which 
leave $u$ unchanged, but we could  think of situations in which 
$u$ has not its stationary value. This corresponds to a 
systems kept far from its ``natural'' configuration.
This can have strong influence even on the critical properties 
of the system as in the case of CA with absorbing states. 
These features will be discussed elsewhere \cite{vz}.

To study the stationary MF solutions, we consider the simple
three level case and we determine $u$ self-consistently
as in Ref.~\cite{vzprl}. Combining the normalization
equation with the stationarity limit of Eqs.~(\ref{mf1})
and \ref{eq:rhos}, we obtain the set of equations
\bea
\nonumber
\rho_a =h\rho_c +(g-\epsilon)\rho_c\rho_a,\\
\nonumber
\rho_a = uh\rho_s + u(g-\epsilon)\rho_s\rho_a,\\
\rho_a =1 - \rho_s - \rho_c.
\label{mf3}
\eea
After some algebra from Eqs.~(\ref{mf3}) 
we obtain a closed equation for $\rho_a$ 
\beq
u(g-\epsilon)\rho_a^2 +(1+u(1+h-g+\epsilon))\rho_a -uh=0.
\eeq
We can expand $\rho_a(h)$ for small values of the field $h$.  
The zero order term in the expansion vanishes and we
obtain a leading linear term:
\beq 
\rho_a(h)=\frac{uh}{1+u(1-g+\epsilon)}.
\label{rhoa}
\eeq
This result has to be consistent with 
the global conservation law, which states that the average
input energy flux $J_{in}$ must balance the dissipated flux $J_{out}$.
In the stationary state the conservation law can be written as
\beq 
J_{in}=hL^d=J_{out}=\epsilon\rho_aL^d.
\label{cons}
\eeq
By comparing Eq.~(\ref{rhoa}) with Eq.~(\ref{cons}) we obtain
that $u=1/(g-1)$, which is the result we have previously
obtained from the complete analysis.
In the limit $h\to 0$ the densities 
are therefore given by
\beq
\rho_a=\frac{h}{\epsilon},~~~~~\rho_c=\frac{1}{g}+{\cal O}(h),
~~~~\rho_s=\frac{g-1}{g}+{\cal O}(h).
\label{mfstasol}
\eeq
The numerical values for the density of critical and stable sites
are non-universal quantities and depend on the lattice geometry
and  dynamical rules of each specific model via the parameter $g$.
The result for $\rho_c$ can be directly compared with the estimates 
from numerical simulations of several models by substituting the 
correct value of $g$. For the original BTW model ($g=2d$), 
extensive numerical 
simulations on the density of energy levels can be found
in Ref~\cite{katori}. As expected the agreement with the MF result 
increases for high dimensional systems and we recover the exact 
result in the limit $d\to\infty$. 

\begin{figure}[htb]
\narrowtext
\centerline{
        \epsfxsize=7.0cm
        \epsfbox{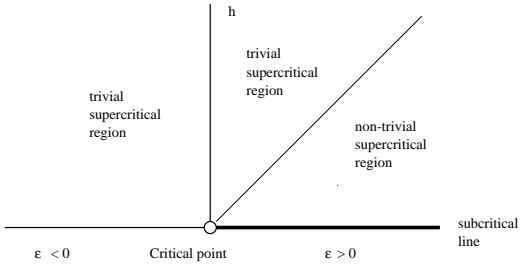}
        \vspace*{0.5cm}
        }
\caption{Phase diagram of a generic sandpile model. We include negative 
values of dissipation corresponding to a net addiction of energy during
the avalanches. The trivial supercritical regime is given by a saturation 
of the system, which is receiving more energy than it can dissipate. 
The interesting region $h<\epsilon$ is discussed in the text.}

\label{fig:1}
\end{figure}

We next discuss the critical behavior of these systems.
The balance between conservation laws and the dissipation are 
essential for the critical behavior of the model, as also pointed out
in \cite{amos}. 
The model is critical just in the 
double limit $h,\epsilon \to 0, h/\epsilon\to 0$, 
similarly to the forest-fire model\cite{dro}.
We are going to see that in this limit the zero field susceptibility 
of the system is singular, signaling a 
long-ranged (critical) response function.
The onset of the critical behavior is then recovered in the 
limit of vanishing driving field corresponding to the locality-breaking 
in the dynamical evolution.    
In analogy with non equilibrium phenomena \cite{torre,jmendes}, 
the one particle density of active sites is the {\em order parameter}
and goes to zero at the critical point. The driving and dissipation rates
identifies the two {\em control parameters}, i.e., the 
relevant scaling fields.
We can then distinguish different regimes as
a function of the control parameters. 
The model is supercritical for
$h>0$ and $\epsilon>h$, while for $h\to 0$ and $\epsilon>0$ 
it is subcritical and the dynamics
displays avalanches. The phase diagram  is somehow similar to
that of usual continuous  phase transitions, if we replace 
$h$ by the magnetic field
and $\epsilon$ by the reduced temperature (see Fig.~\ref{fig:1}). 
We can fully exploit this analogy by allowing the parameter $\epsilon$ to 
assume also negative values. This correspond to a sandpile in which 
a positive net amount of energy enters the system during the avalanche 
activity (negative dissipation). The resulting supercritical 
regime is analogous to many nonequilibrium systems with negative 
reduced control  parameter.
However, for $h>\epsilon$ the system has only trivial stationary state, 
since $\rho_a$ would have to be greater than one to 
satisfy Eq.~(\ref{cons}).
Thus, in the non-trivial supercritical region $h$ and $\epsilon$ can not be 
varied independently because 
the global conservation imposes that $h<\epsilon$. This restricts 
the scaling behavior to particular limit values of the control parameters.
In the following we individuate the regimes corresponding to the standard  
sandpile numerical simulations and a completely 
new scaling regime in the supercritical 
region  of the phase space.

\subsection{The subcritical regime}

The standard numerical simulations of 
sandpile models are carried out in the presence 
of an infinite timescale separation. 
As already discussed in the previous sections, 
this implies an infinite slow driving 
of the systems, i.e. $h\to 0$. In this limit
there is a single control parameter, 
the intrinsic or boundary dissipation $\epsilon$, 
and the order parameter 
$\rho_a$ is identically zero in the steady state.
To describe quantitatively the 
critical behavior, we study the effect of a small 
perturbation $\Delta h$ on the steady state density
\beq
\Delta\rho_a(x,t)=\int\chi_{h,\epsilon}(x-x'; t-t')\Delta h(x',t') dx' dt'
\eeq
where $\chi_{h,\epsilon}(x-x'; t-t')$ is the 
response function of the system.
We define the total susceptibility $\chi_{h,\epsilon}$ 
of the system as the
integral over space and time of the response function, and it is shown in 
appendix A that for the stationary state 
\beq
\chi_\epsilon\equiv\lim_{h\to 0}\chi_{h,\epsilon}=
\left.\frac{\partial \rho_a(h)}{\partial h}\right|_{h=0}.
\eeq
This immediately gives the  zero field susceptibility  
\beq
\chi_\epsilon = \frac{1}{\epsilon},
\label{susc}
\eeq
which diverges as $\epsilon \to 0$. The system is in a 
subcritical state for any value 
of $\epsilon$ different from zero. For $\epsilon=0$ the 
system reaches a critical point in which the response 
function becomes long-ranged and the susceptibility diverges.
Close to this critical point the scaling behavior is 
characterized by the scaling 
laws $\chi_\epsilon \sim \epsilon^{-\gamma}$, 
with $\gamma=1$, and by the divergence 
of the correlation length $\xi\sim \epsilon^{-\nu}$.
 
An important result can be derived  from the response function by defining
\beq
\chi_{\epsilon}(r)=\int \chi_{\epsilon}(r,t) dt
\eeq
as the average total response received at position $r$, 
when $\Delta h(x',t') = \delta(x')$
Since energy is transfered locally and isotropically,
the net energy current is given by $j\sim\partial \chi(r)/\partial r$. 
For locally conservative models 
the energy current $j$ must satisfy in average the conservation law
\beq 
\int j d\sigma=cost,
\eeq 
where $d\sigma$ is 
the $d-1$-dimensional surface element.
This ensures that the energy flowing into the system 
is balanced by the dissipated energy in the stationary state.
Hence, at large $r$ we have the solution $\chi(r)\sim r^{2-d}$.
A similar result has been obtained in Ref.\cite{zhang}.  
In the presence of boundary or intrinsic dissipation, the system 
acquires a finite correlation length and we can 
establish the general scaling form
\beq
\chi_{\epsilon}(r)=\frac{1}{r^{d-2}} ~ \Gamma(r/\xi),
\eeq
where $\Gamma(r/\xi)$ is a cut-off function for $r>>\xi$. 
This immediately gives the following relation between
$\xi$ and the zero field susceptibility
\beq
\chi_\epsilon=\int\chi_{\epsilon}(r)r^{d-1}dr\sim \xi^2
\eeq 
We find the MF value of the correlation exponent 
$\nu=1/2$ by 
substituting $\xi\sim\epsilon^{-\nu}$ and comparing with eq.(\ref{susc}).

We can use these exponents to characterize the finite size scaling of the 
conservative sandpile model, since
our MF analysis treats both boundary and bulk dissipation in the same way.
In conservative systems, when the size is increased  
the effective dissipation depends on the system size
and we assume that $\epsilon\sim L^{-\mu}$. 
In fact, the dissipation rate is given by the probability 
to find a border site instead of a bulk site during 
an avalanche.Thus, the exponent $\mu$ links the dissipation rate 
with the system finite size, providing a unified view of locally 
dissipative and open-boundary models.
In the conservative case, the characteristic length of avalanches 
should go like $\xi\sim L$ to ensure dissipation of 
energy outside the boundaries. 
This implies the scaling relation $\nu\mu=1$, which immediately gives $\mu=2$.
We show in appendix A that the susceptibility scales as the average
avalanche size 
\beq
\chi_\epsilon\sim <s>,
\label{eq:susc} 
\eeq 
which implies $\chi\sim L^{\mu\gamma}$. 
From this result we obtain the scaling law
\beq
\langle s\rangle\sim L^2 \mbox{ for } L\to\infty,
\eeq
which has been found numerically in various dimensions \cite{grasma}
and and rigorously proven in $d=2$ \cite{dhar}. 
Our explanation implies that the
diffusive behavior of the avalanches is due to 
the global conservation law.

Summarizing all these results we obtain a first set of MF exponents
\beq
\gamma=1, ~~~~\mu=2, ~~~~\nu=1/2.
\eeq
In deriving these exponents we made only use of conservation 
laws, therefore we expect that these results hold in
all dimensions. In the next section we will confirm these 
results by numerical simulations performed in the fast 
driving regime and we will discuss some results already 
published in the literature.

In the subcritical regime the dynamics takes place in 
the form of avalanches,
but if $h=0$ the system rapidly 
decays in one of the adsorbing configurations;
the ones with no active sites. All of 
them are stable in absence of the driving field.
It is useful to characterize the proliferation 
of active sites starting from a seed initial 
condition. In close analogy with
CA with adsorbing 
states, we study the spreading of active sites
after a small perturbation. 
We prepare the system in 
an initial state consisting of a 
single active site, i.e. an infinitesimal 
perturbation in the driving field $\Delta h(x,t)=\delta(t)\delta(x)$. 
Since $h(t>0)=0$, active sites cannot be produced 
spontaneously from critical sites and 
can only appear due to the 
spreading of the initial perturbation. 
The properties of this process close to the critical 
point characterize the avalanche behavior typical
of SOC phenomena. 

Following Grassberger and de la Torre \cite{torre} we consider 
the probability that a small perturbation  activates 
$s$ sites (an avalanche in the SOC terminology)
\beq
P(s,\epsilon)=s^{-\tau}{\cal G}(s/s_c(\epsilon)),
\eeq
where $s_c\sim\epsilon^{-1/\sigma}$ is the cutoff in the avalanche size. 
The perturbation decays in the  stationary subcritical state as 
\beq
\rho_a(t) \sim t^{\eta}{\cal F}(t/t_c(\epsilon)).
\eeq
Here $t_c$ denotes the characteristic time which scales as
$t_c \sim \epsilon^{-\Delta}$.
We can also introduce the scaling exponents which relates cutoff lengths
to the characteristic size, $s_c\sim\xi^{D}$ and $t_c\sim\xi^z$.
These exponents are related by the following scaling laws
\beq 
D=\frac{1}{\nu\sigma},~~~z\nu=\Delta.
\label{scal1}
\eeq
Another scaling relation between critical exponents can be  
obtained from Eq.~(\ref{eq:susc}) (see Appendix A)
\beq
\chi_\epsilon\sim <s> = \int s^{-\tau+1}{\cal G}(s/s_c(\epsilon))ds
\sim\epsilon^{(\tau-2)/\sigma},
\eeq
which implies
\beq 
\gamma=\frac{(2-\tau)}{\sigma}.
\eeq

To obtain the MF values of the avalanche exponents, we solve the 
evolution equation for a small perturbation close to the stationary state. 
We consider $\rho_\kappa(t)=\rho_\kappa+\delta\rho_\kappa(t)$, where 
$\delta\rho_\kappa(t)$ is the deviation of the densities from their stationary 
value. By considering small perturbation around the 
stationary state, keeping 
only linear term in $\delta\rho_\kappa(t)$, and 
using the normalization condition we obtain
\beginwide
\bea
\nonumber
\frac{\partial}{\partial t}\delta\rho_a(t)=- \delta\rho_a(t)+
h\delta\rho_c(t)+(g-\epsilon)\rho_c\delta\rho_a(t)
+(g-\epsilon)\rho_a\delta\rho_c(t)\\
\nonumber
\frac{\partial}{\partial t}\delta\rho_s(t)= 
+\delta\rho_a(t)-\frac{h}{g-1}\delta\rho_s(t)+
-\frac{g-\epsilon}{g-1}\rho_s\delta\rho_a(t)
+\frac{g-\epsilon}{g-1}\rho_a\delta\rho_s(t)\\
\delta\rho_a(t)= -\delta\rho_c(t)-\delta\rho_s(t).
\label{dyn1}
\eea
\endwide
In the subcritical regimes ($h\to 0$) we  
only keep in these equations
the leading terms in $\epsilon$. 
Substituting in Eqs.~(\ref{dyn1}) the densities
given by the solution  of the stationary 
equation for $h\to 0$ (i.e. $\rho_a=0$ and 
$\rho_c=1/g$), we finally obtain the evolution equation in 
diagonal form
\beq 
\frac{\partial}{\partial t}\delta\rho_a(t)= 
-\frac{\epsilon}{g}\delta\rho_a(t),
\label{eq:exp}
\eeq
The solution of Eq.~(\ref{eq:exp}) is given by 
\beq
\delta\rho_a(t)\sim\exp(-\epsilon t/g)
\eeq
which implies $\eta=0$. The last equation also
defines the characteristic relaxation time 
for an infinitesimal perturbation to be $t_c=g/\epsilon$, 
yielding $\Delta=1$.
We compute the remaining exponents 
using a further scaling relation, which we derive in appendix B,
\beq
\frac{(\tau-1)}{\nu\sigma}=z.
\label{eq:mfscal}
\eeq
It is worth to remark that Eq.~(\ref{eq:mfscal}) is
valid only in  MF theory. By combining these relations with those of 
Eq.(\ref{scal1}), we get the second set of MF critical exponents:
\beq
z=2,~~~~D=4,~~~~~\tau=3/2,~~~~~\sigma=1/2.
\eeq
It is worthwhile to remark that the numerical value of these exponents 
is the same as in other MF approaches, but their significance is 
completely different, being defined with respect to a 
different scaling field.
All sandpile models with the same dynamical MF equations share 
the same critical exponents and belong to the same
universality class. However, the degree 
of universality is highly overstated, as usually happens 
in MF approaches. In particular, 
the exponents do not depend on the dimensionality $d$.
The exponents describing
avalanche distributions in low dimensional systems, in general,
will not agree with the results of MF theory.
For instance, it is still controversial
if the BTW and the two-state model are in the same universality class. 
While it was believed for some time that this was indeed the case, recently 
large scale numerical simulations questioned this statement \cite{biham}. 

To compute the value of critical exponents
below the upper critical dimension, we have to use 
renormalization group techniques, which allow 
the correct treatment of the scale-free fluctuations 
present at the critical point.
The renormalization group approach to non-equilibrium system  
presents several difficulties which can be overcame by
suitable approximations \cite{ddrg,pvz}. 

\subsection{The supercritical regime}

The supercritical region is characterized by a finite density 
of active sites; i.e. a non-zero order parameter. 
Close to the critical point, the supercritical 
region correspond to the 
parameter range $h<<1$, $\epsilon <<1$ and $h \stackrel{<}{\sim} \epsilon$.
In this regime the order parameter is linear in $h$ 
\beq
\rho_a\sim h^{1/\delta};~~~~~~~~\delta=1,
\eeq
as we obtain from Eq.~(\ref{mfstasol}). The same result has been 
also conjectured in Ref.\cite{grasma}.
This is analogous to the MF results obtained for contact processes and 
other non equilibrium CA \cite{torre,dick,jmendes}, but it is 
in contrast with previous MF approaches for sandpile models
\cite{bakmf,stellamf}, which yielded $\delta=2$.
This latter incorrect result is due to an inconsistency
present in those studies. The scaling is expressed in terms of 
the average energy $\langle z \rangle \equiv \sum_i \rho_i z_i$ which is
treated as an independent control parameter. As we have just 
shown, $\langle z\rangle$ and $h$ {\em are not independent} in the 
stationary state. The stationary probability distribution of heights 
$\rho_i$ is indeed a function of the driving rate. Moreover,
$\langle z \rangle$ can not be considered as the control 
parameter even for $h=0$, since 
it does not determine completely the state of the system: 
the same value of $\langle z \rangle$ 
describes several states corresponding to
different values of densities $\rho_i$. This is a typical
property of CA with multiple absorbing states \cite{jmendes}.
In preparing an  initial condition 
consisting of a localized active region, one 
has considerable freedom to choose the 
initial state (the adsorbing configuration). 
In order to observe the critical properties of the system, 
we should chose one of the ``natural'' 
initial conditions, which can be obtained by 
the dynamical evolution for infinitesimal driving
in the long time limit. In numerical 
simulations this is equivalent to first prepare the 
system in the stationary
state in presence of the timescale separation, 
and then average over the many different realization of the
avalanches.  

The exponent  $\delta$ 
has been defined in previous works in analogy with usual 
continuous phase transitions, where it characterizes the
scaling of the order parameter in presence of an external field 
when the other critical parameters are set to zero. In SOC, however, 
$h$ and $\epsilon$ are not fully independent because 
the of the $\epsilon >h$
condition. Thus in the supercritical regime, 
the scaling of physical quantities is a general homogeneous function
of the following kind
\beq
f(h,\epsilon)=b^\alpha f(b^{y_h}h, b^{y_\epsilon}\epsilon),
\eeq
where $b$ is a scaling factor and $\epsilon$ can not 
be set to zero for finite $h$.
The scaling behavior of the system is therefore 
particularly complex, as in the case
of conventional phase transitions in the 
region where two control 
parameters are different from their critical values. 
We can define a
consistent scaling regime with respect to the 
reduced variable $\phi=h/\epsilon$
in the double limit $h\to 0$ 
and $\epsilon\to 0$ with 
the supplementary conditions that $\epsilon<<\phi<<1$. 
These limits define a 
parameter region which is identified roughly as $\epsilon<<1$ 
and $\epsilon^2<<h<<\epsilon$.
In this region, the MF 
approximations shows that the order parameter
is positive and scales as 
\beq
\rho_a=\phi^{\beta},
\eeq
with $\beta=1$. The exponent $\beta$  characterizes the 
scaling behavior of the order parameter with respect to
the reduced parameter $\phi$
and should not be confused with the exponent $\delta$.
In order to uncover the scaling of the characteristic lengths of the 
system with respect to the parameter $\phi$, we  
study the evolution of small 
perturbations around the stationary state. 
As in the previous section, we denote by 
$\delta\rho_\kappa(t)$ the deviations
from the stationary state and write the dynamical equation 
keeping only the leading terms. We thus neglect in 
Eqs. (\ref{dyn1})  the terms in $h$ and $\epsilon$, and 
keep the terms in $\phi$. 
For this we compute the first order correction in $\phi$
to the values of the stationary densities. These results to be 
$\rho_a=\phi$, $\rho_c=(1-\phi)/g$ and $\rho_s=(g-1)(1-\phi)/g$, 
and by substituting in the dynamical MF equations 
we get
\bea
\nonumber
\frac{\partial}{\partial t}\delta\rho_a(t)= 
-(g-1)\phi\delta\rho_a(t)-g\phi\delta\rho_s(t)\\
\frac{\partial}{\partial t}\delta\rho_s(t)= +\phi\delta\rho_a(t)
-\frac{g}{g-1}\phi\delta\rho_s(t)
\label{dyn2}
\eea
By diagonalizing Eqs.~(\ref{dyn2}), we find the 
eigenvalues 
\beq
\Lambda_{\pm}=-\phi f_{\pm}(g)
\eeq
where  $f_{\pm}(g)$ has always positive real part. 
Both eigenvalues are thus negative and linear in $\phi$ and 
represent the inverse of the relaxation timescale of a 
perturbations around the stationary state.
We have therefore that $t_c \sim \phi^{-\Delta'}$, with $\Delta'=1$,  
implying that the characteristic time scales as in the subcritical regime.  
The solution for the spreading of the density perturbation has the form 
$\delta\rho_a(t) \sim {\cal F}(t/t_c)$, yielding as in the
subcritical regime $\eta'=0$. 

Eqs.~(\ref{dyn2})
describe how a localized perturbation decays in the stationary 
state. As in the previous section,
this decay can be related to the susceptibility 
\beq
\chi_{\phi}\sim\int\delta\rho_a(t)dt\sim \phi^{-\gamma'}
\label{susct}
\eeq
with $\gamma'=1$.
A characteristic length $\xi$
is associated with the characteristic time of fluctuations.
Since energy is transferred 
homogeneously and isotropically, we have that 
$\chi_{\phi}\sim\xi^2$, as in the subcritical regime.
By comparing this relation 
with Eq.~(\ref{susct}), we obtain 
$\xi^2\sim \phi^{-1}$, or $\xi\sim\phi^{-\nu'}$ with $\nu'=1/2$.
We can obtain a clearer picture of this behavior using
the avalanche representation.
The condition that the time between two energy addiction   
(the driving timescale) is much longer than the 
fluctuation timescale can be written as
\beq
h<< t_c^{-1}\sim\phi.
\eeq
This condition is implicitly verified in the limit we are considering
($h<<\epsilon<<\phi$). Under this assumption it is very unlikely
for fluctuations to overlap. Thus, on average, each event is 
separated from the others and can be defined as in the subcritical
case  as an avalanche. 
In the stationary state 
the average non-zero order parameter
is produced by the random appearance of a finite number of non-overlapping
avalanches. Furthermore, we can identify the response function with the time 
evolution of an avalanche, and we recover in the limit $h\to 0$
that $\chi_\phi\sim <s>$. On its turn the latter implies that in this regime
the average sizes of avalanches diverges as $<s>\sim\phi^{-1}$ as 
$\phi\to 0$.

These results are particularly interesting because it is the first time 
that the  supercritical
scaling behavior is characterized by the parameter $\phi=h/\epsilon$.
The MF predictions, also if not exact from a quantitative point of view,
should reflect the right symmetry of the problem. 
Thus we expect that a supercritical 
regime should be found also from numerical 
simulations, but with different 
values of the critical exponents. Although the analytical 
formulation of the  supercritical sandpile is quite simple, the actual 
implementation of the numerical 
simulations is not straightforward since
the scaling regime is obtained 
only in the double limit $h\to 0$ and 
$\epsilon\to 0$. It is not possible to use anymore the timescale 
separation, because we want 
investigate the behavior for finite $\phi$, when 
a single timescale represents the system.
Work is in progress to obtain a numerical confirmation 
of the MF predictions for the supercritical regime.

Usually the equilibrium version of the fluctuation-dissipation theorem allows 
the calculation of the upper critical dimension of models by requiring 
the consistency of the MF theory. The theorem  links the 
susceptibilities 
with the equilibrium fluctuations from which Ginzburg-like 
self-concistency criteria are derived.
The existence of similar theorems for nonequilibrium steady-states would be 
clearly useful. From general considerations, it is possible to show 
that fluctuation-dissipation-like  theorems do exist for nonequilibrium
models. However, the physical quantities that satisfy these theorems 
are generally unknown, and their construction requires the complete 
solution of the system's dynamics. This fact is very unsatisfactory from a 
practical point of view and has given rise the terminology of 
``violation of the fluctuation-dissipation theorem''. For stochastic 
SOC we are therefore unable to use the susceptibility in order to estimate 
the stationary fluctuations. This does not allow, at the present level 
of approximation, to find the upper critical dimension $d_c$ of this class of 
models. From the theoretical  point of view the situation is still very 
controversial. Several theoretical estimates give $d_c=4$ \cite{diaz}, that 
has been obtained also from recent numerical simulations
\cite{lubeck}.
In contrast, other numerical studies\cite{co} and 
the similarity to percolation
led several authors to conjecture $d_c=6$.
Despite the fact that exponents
have the same numerical values, our approach points out that  
the MF theory for SOC models is very different from that
of percolation. For instance, we have shown 
that the exponent $\delta$ is not well defined in our case. 
We can understand therefore that
some scaling relations such as $\delta=(\tau-2)^{-1}$
which are valid for percolation do not apply here.

\subsection{Numerical simulations}

In this section we compare the results of numerical simulations
with the prediction of MF theory, and in particular 
we expect that the set of MF exponents related to the global conservative 
nature of the model are valid  also in the low dimensional cases. 
Sandpile models have been 
extensively studied only in the subcritical 
state \cite{kada,grasma,manna1,lubeck}. 
Most of the numerical results refer to avalanche distribution in the
conservative limit, with open boundary conditions. In these
conditions the finite size scaling has been found to be
 problematic and despite
the use of very large scale simulations there is not a complete
agreement on the values of the exponents \cite{lubeck}. The reason for this
is probably that open boundary conditions
impose a value for the effective dissipation which 
depends on the lattice size and does not act homogeneously
through the system.

\begin{figure}[htb]
\narrowtext
\centerline{
        \epsfxsize=7.0cm
        \epsfbox{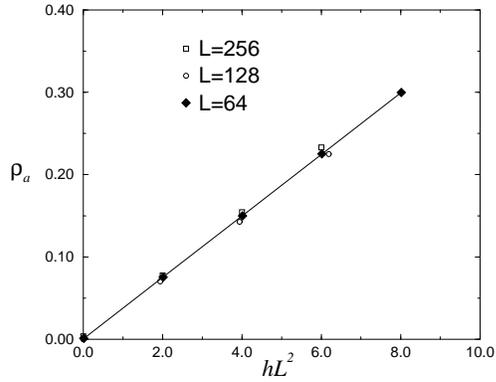}
        }
\caption{The density of active site in the BTW model
with boundary dissipation as a function of the driving rate $h$
is plotted for different system sizes $L$.}
\label{fig:2}
\end{figure}

We simulate numerically the BTW model with finite driving rate 
$h$ and boundary dissipation in $d=2$. In this case the dissipation
is implicitly considered through the open boundary conditions. When
a boundary site topples, it dissipate part of the energy outside, without
transferring it to the neighbors. The driving rate $h$ is introduced as the
probability for unit time that a site receives an energy grain. Apart
from the driving the simulation proceeds as in Ref.~\cite{grasma}. 
We see in Fig.~\ref{fig:2}
that the density of critical sites goes to zero linearly with
$h$ ($\delta=1$) with a slope that increases with the system size 
as $L^{2}$. This is in agreement with the MF theory which
predicts that the susceptibility scales as $L^{\mu\gamma}$,
with $\mu\gamma=2$.

\begin{figure}[htb]
\narrowtext
\centerline{
        \epsfxsize=7.0cm
        \epsfbox{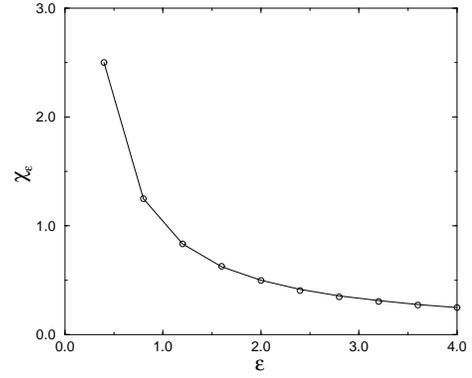}
        }
\caption{The susceptibility $\chi_\epsilon = d\rho_a / d h$ as a function
of the dissipation $\epsilon$, for a system with periodic boundary
conditions and size L=64. The line corresponds to the theoretical prediction
$\chi_\epsilon = 1/\epsilon$.}
\label{fig:3}
\end{figure}

To observe more clearly the scaling with dissipation 
of the sandpile model we study the BTW model 
with {\em periodic} boundary
conditions and fixed dissipation $\epsilon$. 
We model the dissipation introducing  a probability 
$\tilde{p}\equiv\epsilon/g$,
for which the energy in a relaxation event is lost, 
instead of being transfered. 
In Fig.~\ref{fig:3} we plot the susceptibility 
$\chi_\epsilon=d\rho/dh$ as a function of $\epsilon$.
We observe the $1/\epsilon$ behavior ($\gamma=1$)
predicted by mean-field theory. 
We should add to this the value of $\nu = 0.5 $ that was
obtained studying a dissipative sandpile model in $d=2$ \cite{diss}. 

In summary, we have shown that some MF features are present in the 
$d=2$ sandpile model. The global conservation law imposes the exponents 
$\mu,\nu$ and $\gamma$ to assume their MF values. 
This strongly support the MF picture provided here.
The other critical exponents do not necessarily  assume mean-field values
in $d=2$ and we do not analyze them here. Extensive 
measures of these exponents can be found in the literature 
\cite{kada,grasma,manna1,lubeck}. We are currently 
performing simulations of the model
in the supercritical regime
in $d>2$ and the results will be published elsewhere \cite{chessa}.

We expect the critical properties, such as 
exponents and scaling, to be universal for the 
class of ``natural'' configurations.   
It is  interesting to note that spreading exponents 
could be sensitive to initial configurations 
different from the ``natural'' ones. This has been 
extensively studied in nonequilibrium phase transitions, where 
the critical behavior is changing with respect to the adsorbing 
configurations used as initial conditions. In equilibrium critical phenomena, 
the appearance of continuously variable exponents is usually 
associated with a marginal parameter, that can be the case for critical 
phenomena at adsorbing states. 
Unfortunately, no numerical data are available on sandpile models 
to place these speculations on a firmer basis. 

\subsection{On the role of conservation}

In the previous discussion, we emphasize the role of 
conservation in the dynamics of sandpile models. Using
conservation laws, we found a subset of critical exponents
that retain their mean-field values even in low dimensions.
Conservation has an important effect on criticality as well,
since the amount of dissipation plays the role of control 
parameter. Only by tuning this parameter to zero the system
is critical.

The role of conservation in SOC models has been
the object of a long controversy. It was first claimed
that conservation was a necessary conditions for criticality
in this class of models \cite{hwa,grin2}. 
The result on dissipative sandpile models seemed to confirm
this conclusion \cite{diss}. 
Later on, simulation of an earthquake model contradicted this
results \cite{ofc}. The model studied in \cite{ofc} is a 
continuous height sandpile model in which energy in dropped
uniformly over all the lattice at infinitesimal rate.
This in practice corresponds to rise the heights of all the sites
by the quantity needed for the higher site to become unstable.
When a site $i$ is unstable ($z_i>z_c$) the relaxation rules are
\beq
z_i \to 0
\eeq
\beq
z_j\to z_j+\alpha z_i
\eeq
where $j$ are the nearest neighbors of $i$. The dissipation
parameter $\alpha$ \cite{nota_alpha} can be tuned: for $\alpha =1/(2d)$ the
system is conservative. It has been observed in simulations
that in $d=2$ for $\alpha > \alpha_c$ the system is critical.
There is still not an agreement on the precise value of $\alpha_c$,
which was first estimated as $\alpha_c \simeq 0.05$ \cite{ofc} 
and later found to be higher ($\alpha_c \simeq 0.18$) \cite{ofc2},
while it was claimed in Ref.~\cite{mt} that $\alpha_c=0$.

The mean-field analysis of this model is not 
easy because of the continuous number of levels a site
can assume. Using an approximate analysis of the random-neighbor
model it was claimed in \cite{lise} that $\alpha_c \simeq 0.22$,
a value that was found in agreement with simulations.
A complete analysis of the master equation later revealed that
$\alpha_c=0.25$ (the conservative case), showing also the presence
of very strong finite size correction \cite{rnofc}. 
From this analysis, it appears that the random neighbor model
behaves like the BTW model. Criticality is only reached in 
the conservative case, in the limit of zero driving rate.
The situation in two dimensions is still controversial
and it is believed that the inhomogeneity created by the 
open boundary conditions is
responsible for the observed power law distributions.

The role of conservation for criticality remains still
open in this models, while it is now agreed that in MF
theory conservation is a necessary condition to achieve
criticality in sandpile models. In the next section
we will discuss a model \cite{dro} that shows
criticality without conservation even in MF theory. The
price to achieve this result will be an additional driving rate.

\section{NON CONSERVATION AND CRITICALITY: THE FOREST-FIRE MODEL}

We have discussed that conservation in sandpiles is crucial
to achieve criticality.
The controversial issue of continuously driven models 
raises the question of the possibility that timescale separation alone 
can produce scale invariance in systems without conservation laws. 
In this context the forest-fire  model acquires 
a very important role in that it is a nonconservative 
automaton displaying criticality. 

As outlined in Sec.III we can describe the model in the same
language used for sandpile models  
We identify burning sites with the active sites
since they interact with other sites 
independently of the driving fields. Furthermore, their density 
vanishes in the limit of small driving field $f$.
In the same way, trees correspond with 
critical sites and empty sites with stable sites. In this case 
the general three state description 
is exact. Using this language, we can emphasize 
differences and analogies between 
FF and sandpile models. While the main dynamical 
transitions are very similar, we can immediately recognize the effect of 
non conservation. In the FF model energy in not stored 
and new critical sites are created 
by a second independent field, the tree growth probability $p$.
Thus in the FF model we replace $h\to f$ and $uh\to p$. The last 
substitution introduces a new independent field 
(or a new timescale $p^{-1}$)
related to the injection of critical sites in the system. 
Since energy is not accumulated, there is no need of an
additional dissipation, so that in the FF model there is no
parameter playing the role of $\epsilon$.

The FF model was originally introduced in the limit $f=0$ and $p\to 0$
by Bak, Chen and Tang in Ref.\cite{bct}. The model was claimed to show SOC, 
but later Grassberger and Kantz \cite{graka} 
showed that in $d=2$ the model was critical 
in a trivial sense. The system shows a diverging characteristic 
length that is essentially the distance between straight fire fronts. 
This implies that the dynamics is governed by the average tree 
density over larger and larger regions. In higher dimensions the 
possibility of a non trivial behavior has not been ruled out as recent work 
seems to suggest \cite{bro}.
Drossel and Schwabl \cite{drosch} introduced the ignition 
or lightning probability $f$. This field sustain fires and the 
system flows in a stationary state which shows critical properties 
in the double limit $f<<p<<1$. 
This version of the model has been the subject of several studies both 
analytical \cite{ddrg,mfff,drocla,teoriaff,teoriaff2} 
and numerical\cite{clar,hen,grasff}.

Despite the various efforts, the two versions of the model were 
always studied as very different cases, almost two different models. 
For this reason, it is difficult to find in the literature 
a precise connection among the 
two different regimes. In this section we recover 
within our framework  many results already present in the literature. 
By recasting these results in the language developed for sandpile automata,
we provide a unified picture of both models.
We discuss the FF model in terms of the response function singularities
and we show
that the SOC-FF and the deterministic FF correspond to the supercritical 
and subcritical regimes.
In this way, we can understand  several features of the FF model in terms
of the same concepts developed for sandpile model.

The MF equations can be derived by the 
single site approximation to the master equation.
Since the derivation proceeds as in section IV,  we do not repeat it
here in detail and we present instead some general considerations.
Active sites become stable (fire $\to$ empty) 
with unitary rate, and critical sites 
become active if ignited by the lightning with probability $f$. 
The interaction term is then given by the fire spreading; an active site 
create as many new active  sites as the number of  n.n. critical sites. 
To the first order in $\rho_a$, this term is proportional to $\rho_a\rho_c$
times the usual geometrical factor $g$ that takes into account the lattice 
coordination and other model dependent geometrical effects.
The reaction rate equations then reads
\beq
F_a= -\rho_a +f \rho_c + g\rho_c\rho_a +{\cal O}(\rho_a^2).
\eeq
This expression is very similar to the one obtained in the sandpile case,
with the exception of the dissipative term that here is missing.
The two models differ in the dynamical evolution 
of stable sites. In the FF model there is a
term, due of the field $p$, corresponding to the transition 
rate from stable to critical sites 
and there is no interaction between active and stable sites. 
We can then write the stationarity equations
for the FF model as
\bea
\nonumber
\rho_a =f\rho_c +g\rho_c\rho_a\\
\nonumber
\rho_a = p\rho_s\\ 
\rho_a =1 - \rho_s - \rho_c ,
\label{mffor}
\eea
where we have neglected second order term in $\rho_a$.
As for the sandpile model, $g$ is an independent parameter of the model and 
$f$ and $p$ represent the tunable external driving fields. 
The lowest order solutions in $f$ and $p$ to the above equations are 
\bea 
\rho_a&=\frac{g-1}{g}p + \frac{1}{g}f + {\cal O}(f^2,p^2)\nonumber\\
\rho_c&=\frac{1}{g} - \frac{1}{g}\frac{f}{p} + {\cal O}(p,f)\nonumber\\
\rho_s&=\frac{g-1}{g} + \frac{1}{g}\frac{f}{p} + {\cal O}(p,f). 
\label{ffsol}
\eea
These results have been already obtained in Ref.~\cite{mfff}, where 
a random neighbor version of the FF is analyzed. Their method and the 
present MF scheme are equivalent and we will recover the same stationary densities. 
We compute the critical exponents by using the same 
lines adopted for sandpiles, and obtain some new insight on the critical 
properties of the FF model. The density of active sites depend 
linearly upon $f$ and $p$, which are independent driving fields 
playing the same role as $h$ in sandpile automata.
If we consider the density of active sites as the order parameter, it appears 
immediately that the critical point is reached if $f\to 0$ and $p\to 0$
simultaneously. This double limit again corresponds 
to the locality breaking of the dynamical rules.
In this case the order parameter is identically zero in 
the steady  state and the system develop long range correlation properties.
Also for the FF model we can then distinguish among a 
subcritical and a supercritical regime depending on the values of 
the driving fields. 

\subsection{The subcritical regime}

The subcritical and critical regimes correspond to the limit 
in which we have zero order parameter and therefore $f=0$ and $p=0$.
This limit is however not completely
defined because the density of critical and stable 
sites depend upon the ratio $f/p$. In order to study the critical behavior 
in this limit we repeat the discussion inspired by the study of 
CA with adsorbing states that we already used for the sandpiles. 
Since when $f=0$ and $p=0$ the dynamics is frozen, we have to 
prepare the system in a stationary state in the limit $p\to 0$ and $f\to 0$, 
and then study the spreading of small perturbations.
This is what is actually done in numerical simulations 
where the fire evolution and the action of $f$ and $p$ act separately.
In doing that however, we prepare the system in one of the ``natural'' 
configurations, corresponding to the stationary state 
in the limit of infinitesimal driving. In this configurations the density of 
critical sites reaches a limit value $\rho_c= 1/g -g^{-1}f/p$, which 
depends, via the parameter $\theta\equiv f/p$,
on the way the limit has been performed. 
In order to study 
the scaling behavior, we consider the limit  $f<<p<<\theta<<1$, keeping 
$\theta$ constant. In this regime we can consider 
$f=p=0$, $\rho_a=0$ and we can study the system to the leading 
order in $\theta$. 
By considering  small deviations $\delta\rho_{\kappa}(t)$
from the stationary state and retaining just first order terms in $\theta$,
we find the linearized dynamical equation in diagonal form
\beq 
\frac{\partial}{\partial t}\delta\rho_a(t)= 
-\theta\delta\rho_a(t).
\eeq
Hence, the relaxation behavior follows an exponential law 
in which the characteristic relaxation time is given by 
$t_c\sim \theta^{-\Delta}$ with $\Delta=1$. 
This implicitly tells us that the system indeed reacts 
in avalanches. In fact, both driving timescales $p^{-1}$ and 
$f^{-1}$ are in this regime much longer than the characteristic spreading 
time of an avalanche $\theta^{-1}$, that therefore remains an 
isolated event connected in space and time.

Along the lines we followed for sandpiles, we 
define the response function of the system 
$\chi_{f,\theta}(x-x'; t-t')$ that characterizes the way the system
respond to an external perturbation. The response is now function of 
$f$ and $\theta$ which is fixed. The total susceptibility $\chi_{f,\theta}$
is related to the derivative of the stationary density of critical sites, 
and the zero field susceptibility can be obtained as
\beq
\chi_{\theta}=\lim_{f\to 0}\frac{\partial \rho_a(f)}{\partial f}.
\eeq
Since the density of active sites can be written as
$\rho_a(f,\theta)=f/g +(g-1)f/(g\theta)$, the 
singular part of the susceptibility diverges as
\beq
\chi_{\theta}=\theta^{-1}.
\label{ffre}
\eeq
In appendix A it is shown that the zero field susceptibility is related 
to the divergence of the average fire size as $<s>\sim\chi_{\theta}$. 
Hence, the characteristic fire size is diverging for $\theta\to 0$.
This implies that the system is in 
a subcritical regime and perturbations to the stationary state
show a finite characteristic length for any $\theta > 0$. 
Only in the limit $\theta\to 0$ the system responds
on all length scales to infinitesimal 
perturbations. We can define the standard scaling laws 
$\chi_{\theta}=\theta^{-\gamma}$ with $\gamma=1$,
and $\xi\sim\theta^{-\nu}$ that characterize the divergence of 
the correlation length. 

Next, we consider the total response at position
$r$ given by $\chi_{\epsilon}(r)=\int \chi_{\epsilon}(r,t) dt$. 
We note that fire clusters are given by the 
connected clusters of critical sites, because in this regime fires are not
overlapping. Since a tree can burn just once, the average response at
distance $r$ is the given by the pair connectedness function which is 
supposed to behave as $r^{d-2}~\Gamma(r/\xi)$ 
in MF theory \cite{conn}. In general, by integrating the local 
response function, we have
\beq 
\chi_{\theta}=\xi^2, 
\eeq
and therefore by comparing with Eq.~(\ref{ffre}) we get $\nu=1/2$.
It is worth to remark that in this case the above MF relations are not 
enforced by conservation laws and anomalous 
exponents can appear in low dimensions.

To study the avalanche behavior,  we introduce the probability
$P(s,\theta)=s^{-\tau}{\cal G}(s/s_c(\theta))$
that a fire involves $s$ sites and we identify the usual set of critical 
exponents defined by the scaling laws $s_c\sim\theta^{-1/\sigma}$, 
$s_c\sim\xi^D$ and $t_c\sim \xi^z$. Associated to them we have the 
scaling relations  $D\sigma=1/\nu$ and $\gamma\sigma=2-\tau$.
We have shown previously that in this regime the class 
of ``natural'' configurations have a density of critical sites which 
depends on $\theta$, thus
we consider the difference of densities 
with respect to the critical state 
\beq 
\rho_c -\rho_c(\theta)\sim \theta^\zeta~~~~~~~~~~~~\zeta=1.
\eeq
We can then 
find another scaling relation that links the avalanche exponents to
$\zeta$, noting that 
the avalanche size distribution 
corresponds to the distribution of connected critical sites cluster.
In appendix B we derive this scaling relation which results to be 
\beq
\zeta=\frac{\tau -1}{\sigma}.
\eeq

Collecting all the results obtained above, we have the complete 
set of MF exponents:
\beq
\gamma=1, ~~~~~~~~~~~~\nu=1/2,\nonumber
\eeq
\beq 
\tau=3/2,~~~~~~D=4,~~~~~~~ \sigma=1/2~~~~~~~z=2.
\eeq
Also in this case, as previously shown by several authors\cite{mfff},
the MF values correspond to those of mean-field  percolation.
It is important to stress again that in the FF case, the absence of 
a conservation constraint implies that MF values for critical exponents
are not valid in low dimensional system. Anomalous scaling appears below
the upper critical dimension and the model shows non-trivial values of 
exponents \cite{dro}. 

\subsection{The supercritical regime}

We consider here the scaling behavior in the region 
in which the order parameter is not zero.
In order to remain in the critical region,  we must have $\theta<<1$,
but now we consider non vanishing $f$ and $p$, with $f$ much 
smaller than $p$.
This essentially corresponds to the FF model without ignition that in 
this perspective can be considered as the supercritical regime close to the 
critical point. In this limit we obtain immediately from 
the solution of Eq.~(\ref{ffsol}) that the order parameter 
is positive and scales as 
\beq
\rho_a\sim p^{\beta},
\eeq
with $\beta=1$. To calculate the 
relaxation properties we have to perform a linear 
stability analysis of the system around  
the stationary solutions (\ref{ffsol}), retaining only  
the lowest order terms in $p$. We consider small fluctuations 
$\delta\rho_{\kappa}(t)$ and the eigenvalues of the diagonal
form of the dynamical evolutions are
\beq
\Lambda_{\pm}=-(g/2)p\pm i(gp)^{1/2}.
\eeq
The negative real  part identifies the 
characteristic relaxation time that scales as  $t_c\sim p^{-1}$.
Together with the exponential relaxation, the system shows oscillations
with period $T\sim p^{-1/2}$,
related to the imaginary part of the eigenvalues.
This MF behavior has been already discussed in Ref.\cite{graka}. 

In the supercritical state
the timescale of a perturbation is comparable to the driving scale, 
both being of the order of $p^{-1}$. Thus, active sites do not spread 
just on connected cluster of critical sites. In other words the 
critical sites configuration is not frozen during the perturbation and 
the time evolution connects several clusters of critical sites because 
connecting sites might appear during the time evolution. 
Also in this case, though, the susceptibility is given by the 
total response to a localized fluctuation 
\beq
\chi_p\sim\int\delta\rho_a(t)dt\sim p^{-1}.
\eeq
Since the response of the system is due to the connectivity properties, 
we have still the usual MF relation $\chi_p\sim\xi^2$,
which implies that $\xi\sim p^{-\nu'}$ with $\nu'=1/2$. Another way to see
this result is to think that fluctuation spreads as waves of active sites.
Since the propagation velocity is finite, the correlation length
is proportional to the wave period $T$. This simple MF picture does
not work in low dimensions \cite{bro}.

\section{DISCUSSION AND OPEN QUESTIONS}
\subsection{Relations with Branching processes}
A clear mean-field description of the avalanches
in SOC models has been obtained through the mapping to
branching processes \cite{alstr,gp,sobp,sobp2}. 
A branching process \cite{harris} is defined by a number
of active sites that can either die or generate $n$ new 
sites with certain probabilities. 
The simpler example is the case $n=2$: a site dies with
probability $1-q$ or generates two new sites with probability $q$.
The process usually starts with a single active site
and continues until no more active sites are present.
Depending on the value of $q$ the branching process will
die after a finite number of steps or continue forever.
There is a critical value $q=q_c$ that separates the two regimes
($q_c=1/2$ for $n=2$). For $q <q_c$, the size distribution
of the branching process
is a power law 
\beq
P(s) \sim s^{-3/2}f(s/s_0), 
\eeq
where the cutoff $s_0$ diverges for $q=q_c$. 

It has been shown in Ref.~\cite{sobp} that the Manna
model can be exactly mapped into a branching process
with a time dependent parameter $q(t)$ depending on the density
of critical sites ($\rho_c$) and on the dissipation \cite{sobp2}.
A critical branching process was obtained as a stationary state
in the limit of slow driving and conservation \cite{sobp}.

Branching processes can be considered as a general
framework to describe avalanches in mean-field theory. 
In general terms, we can describe an avalanche by
an evolving front that can either propagate or stop.
In the mean-field description, the elements of the
fronts do not interact and evolve independently.
Thus the avalanche can be described as a branching
process with an effective parameter $q$ that depends
on the detail of the model under study.

In our formalism, a branching process is associated with
the propagation of active sites in the subcritical regime.
In the stationary state for $h=0$, an active site
generates $k=1, \dots g$ new active sites with probabilities
\beq
q_k = (1-\epsilon)\left(\matrix{g\cr k} \right)\rho_c^k(1-\rho_c)^{g-k},
\eeq
while no active sites are generated with probability
\beq
q_0 = \epsilon + (1-\epsilon)(1-\rho_c)^g.
\eeq
In this case, the control parameter for the branching
process is given by $\tilde{q} = \sum_k k q_k $ with a
critical value $\tilde{q}_c=1$. In the stationary state, we find
$\rho_c=1/g$ and hence $\tilde{q}=1-\epsilon$. The critical
branching process corresponds therefore to the limit $\epsilon \to 0$.
A similar analysis can be done for the FF model.

\subsection{Locality Breaking}
We have seen that criticality in stochastic SOC systems is achieved 
only in the limit of infinitesimal driving corresponding
to the locality breaking of the dynamical rules. The non-locality 
is evident if we consider the zero driving limit 
that is naturally implemented in computer simulations 
using two different timescales, 
one for the avalanche evolution and one 
for the driving. With this infinite timescale
separation we introduce new perturbations only when 
the system is quiescent; i.e. the evolution of each site  
depends on the entire system. 
For a more concrete physical explanation of how the locality breaking
generates long-range interactions in the system, let us
consider the case of a vanishing driving rate, 
corresponding to a small density of 
active sites. Because of the infinitesimal  
driving each region devoid of active particle is virtually frozen until an 
active site is generated. The activity spreads 
and in general alter the configuration before it moves away or disappears. 
The active sites leave a trace of their dynamical history 
in the frozen configurations of critical and stable sites they produced. 
If new active sites are created in the same region at some 
later times, they can feel the effect of the active sites
present earlier in the region. This is basically a memory effect, which 
creates a long-range interaction in time and space among diffusing 
active sites. The range of this interaction  depends
on the characteristic timescale of the driving, because the 
fluctuations induced by the driving destroy the memory effect. 
Close to the infinite timescale separation, the characteristic 
driving timescale is diverging and the range of the 
nonlocal interaction extends to the entire system. 
A local interaction is recovered, however, if we introduce 
a size cut-off in the wandering region of active particles.
This is the case of dissipative sandpile in which after a finite number of 
steps the active sites disappear\cite{steps}. 
Over this characteristic size active particles do not interact 
and to obtain a long-range non-local interaction 
the dissipation should go to zero. 
The same discussion applies to the FF model, due to the finite range of 
connected critical sites obtained by tuning the ratio of $f$ and $p$. 

In this framework, the infinite timescale separation 
introduces a 
nontrivial long-range interactions between 
active sites which lead
to a singular behavior; i.e. critical properties in time and space.
In sandpile and FF model, dissipations or other driving rates
introduce a finite cut-off to this non-local interactions, which
induces a subcritical regime. The critical point is reached
just in correspondence of a second limit corresponding to the 
celebrated {\em double timescale separation}. 
This framework leaves room for the appearance of systems in which just 
the single timescale separation could be enough to get criticality. These 
system might show a phase diagram without subcritical phase. 

\subsection{Conclusions}
 
In this paper we have presented a unified mean-field theory for
stochastic SOC models. We have treated these model in analogy
with other non-equilibrium cellular automata, using a single-site
approximation to the master equation. With the present approach,
we are able to identify the order parameter and the control parameters
of the models and to emphasize similarity and differences
between SOC and other non-equilibrium system.
In particular, the language of cellular automata with 
absorbing state can be employed to describe
SOC models. For finite driving rates, we find a  
supercritical regime characterized by a finite fraction
of active sites. In the limit of infinitesimal
driving, the system is subcritical and displays avalanche 
response. Criticality arise from a double limit: the driving rate
and the dissipation (in the sandpile model) or the two driving rates
(in the FF model) should have vanishing values. This limit
correspond to the onset of non-local dynamical rules, which
are responsible for the critical behavior characteristic of SOC.

In this perspective, SOC models appear to be a subset of
non-equilibrium system with steady states, reaching
criticality by the fine tuning of control parameters.
While this statement is technically correct, we note that
SOC system are quite peculiar, since the fine tuning can 
only be achieved by limit procedure. This is in contrast
with ordinary critical phenomena, where the control parameter
can be directly tuned to its critical value.
In this sense, SOC systems are less sensitive to fine tuning \cite{grin}.
Moreover, the driving rate can in general be small in many
natural phenomena, and this could make the SOC framework
relevant.  

\section*{ACKNOWLEDGMENTS}
We thank G. Caldarelli, A. Chessa, K. B. Lauritsen, V. Loreto, 
E. Marinari, A. Maritan, M.A. Mu\~noz, A. Omar Garcia and L. Pietronero.
We are grateful to R. Dickman for valuable comments and suggestions. 
A. V. is indebted to J. M. J. van Leeuwen for interesting discussions.
We thank the Instituut-Lorentz in Leiden where part of this work has
been completed. The Center for Polymer Studies is supported by NSF. 

\appendix
\section{Response Function Properties}

For small perturbations around the stationary state, the 
spontaneous microscopic dynamics can be represented by introducing 
the {\em response function}. We first consider 
the sandpile case.  
If we apply a time-dependent perturbation $h(x,t)$ to the
stationary state, the density of active sites changes as 
\bea
\nonumber
\Delta\rho_a(x,t)=\\
\int \int\chi_{h,\epsilon}
(x-x'; t-t')\Delta h(x',t') d^dx' dt' +
{\cal O}((\Delta h)^2)
\eea
where $\chi_{h,\epsilon}(x-x'; t-t')$ is the response  or 
generalized susceptibility function. 
Here we assume a stationary and homogeneous system, 
i.e. the two point averages depend just on the time or space displacement.  
The above expression  is valid in the linear 
regime, only for small variations of the perturbing field. 
We next derive some simple properties of the response function for 
systems whose dynamics is characterized by avalanches. 
We first consider an impulsive disturbance $\Delta h(x',t')=
\delta(t)\delta^d(x)$. This is a very small perturbation with respect to the 
total energy input $J=\int h(x)d^dx$. In practice it corresponds to the 
addiction of a energy grain on top of the stationary average 
driving field. Inserting this perturbation into $\Delta\rho_a(x,t)$ yields 
\beq
\Delta\rho_a(x,t)=\chi_{h,\epsilon}(x;t).
\eeq
We ten define the total susceptibility of the system
\beq
\chi_{h,\epsilon}=\int dt\int \chi_{h,\epsilon}(x;t)d^dx,
\eeq
which quantifies the total response of the system to an impulsive 
disturbance. The total number of active sites 
due to the perturbation is 
\beq 
N_a=\int dt\int\Delta\rho_a(x,t)d^dx=\chi_{h,\epsilon}.
\eeq
In absence of external field $h\to 0$, the only active sites present 
in the system are due to the delta-perturbation. That is, all the active sites 
are casually connected in space and time, thus forming an avalanche whose 
average size is $<s>=N_a$. This is precisely stated by the following 
expression
\beq
\chi_{\epsilon}\equiv\lim_{h\to 0}\chi_{h,\epsilon}=<s>,
\eeq
which defines a relation between the average avalanche size and the 
zero field susceptibility. As we have seen in the previous sections
the above expression is at the basis of several scaling relations 
and it explains together with conservation the diffusive behavior 
of the average activity. 

Another way to look at a stationary perturbation or equivalently to the 
variation of the stationary averages is the following. 
We consider a different perturbation 
\beq
\Delta h(x',t')=\Delta h~~~~~~~~~~~~~~\mbox{for}~~ t'<t,
\eeq
corresponding to a uniform driving in space and time.
By changing the variables of integration to $t''=t-t'$ and $x''=x-x'$ 
we obtain
\beq
\Delta\rho_a=\Delta h\int_V\int_{0}^{\infty}
\chi_{h,\epsilon}(x'';t'')d^dx''dt''.
\eeq
Hence, the density fluctuations are time and space independent, as it must
be in the new stationary state with $h\to h+\Delta h$. Performing the 
double integral in the right term we get the total susceptibility. 
Therefore 
\beq
\Delta\rho_a=\Delta h\chi_{h,\epsilon},
\eeq
from which we obtain that in the stationary state and for 
infinitesimal perturbations 
\beq
\chi_{h,\epsilon}=\lim_{\Delta h\to 0}\frac{\Delta\rho_a}{\Delta h}=
\frac{\partial \rho_a(h)}{\partial h}.
\eeq
Notice that from this equation we are able to provide a relation
between the total response function and the divergence of avalanche 
size 
\beq
<s>=\chi_{\epsilon}=\lim_{h\to 0}\frac{\partial \rho_a(h)}{\partial h}.
\label{eq:smedio}
\eeq
Eq.~\ref{eq:smedio} states that the zero-field susceptibility in the 
stationary state and the average 
avalanche size have the same singular behavior in the thermodynamic limit.

We next consider the Forest-Fire model. In this case we 
have the two driving fields $f$ and $p$ and the response function depends
upon them. The interesting subcritical regime is the one in which we take 
the limit $f\to 0$ and $p\to 0$ with $\theta=f/p <<1$. We study 
the response of the system for small perturbation $\Delta f$ and a fixed
value of  $\theta$. The general expression that characterizes the 
response of the system is given by
\bea
\Delta\rho_a(x,t)= \nonumber\\
\int \!\!\!\int\chi_{f,\theta}
(x-x'; t-t')\Delta f(x',t') d^dx' dt' +
{\cal O}((\Delta f)^2).
\eea
As for the the sandpile case we can apply a delta-perturbation. It 
follows that, simply rewriting what we derived in the sandpile case, we 
obtain
\beq 
\chi_{\theta}\equiv\lim_{f\to 0}\chi_{f,\theta}=<s>,
\eeq
where $<s>$ in this case is the average size of fire events.
In the same way we can consider a stationary perturbation 
$\Delta f(x',t')=\Delta f$ for $t'<t$ and by repeating the above arguments 
we recover
\beq 
\chi_{f,\theta}=\frac{\partial \rho_a(f)}{\partial f},
\eeq
from which follows that the divergence of the average size of 
fires is related to the zero field susceptibility in the usual way. 

\section{Scaling Relations}

Here we obtain two scaling relations whose derivation 
is straightforward but rather lengthy. 

\subsection{Sandpile model}

Let us consider the flow decays of activity in the subcritical 
regime. We define $\rho_a(s,t)$ the space integrated response of an avalanche
of $s$ sites. If we assume scaling behavior we have that 
\beq
\rho_a(s,t)\simeq s^q w(t/t_s)
\eeq
where $t_s$ is the upper characteristic time of an avalanche of $s$ sites
and scales as $t_s\sim s^{z/D}$. By imposing the condition that 
$\int \rho_a(s,t)dt=s$ we obtain $q=1-z/D$. We have also that $w(0)=w(1)=0$
and that $\rho_a(s,t)$ is independent of $s$ for small $t$. This 
implies that $w(x)\to x^{-1+D/z}$ as $x\to 0$.
The total response function is the average of the various possible avalanche 
response 
\beq
\rho_a(t)=\int\rho_a(s,t)P(s)ds
\eeq
that gives after the proper substitution by means of scaling
relations  the expression:
\beq
\rho_a(t)\sim \exp[ -(t/t_c)^{(\tau-1)/\nu\sigma z}].
\eeq
In the MF picture the above relation is consistent with the results 
obtained from the dynamical equations only if 
\beq
\frac{(\tau-1)}{\nu\sigma z}=1
\eeq
thus recovering the relations used in Sec. IV.  An analogous results 
was already obtained in the paper by Tang and Bak \cite{bakmf}. 
Again we stress that this is not a general scaling relations, 
but an exponents equality valid just in MF theory. 

\subsection*{Forest-fire model}
The density of critical sites in the stationary  configuration
approaches the critical value for $\theta\to 0$ as a power law 
\beq
\Delta\rho_c\equiv\rho_c(\theta=0)-\rho_c(\theta)\sim \theta^\zeta.
\label{ffapp}
\eeq
In the subcritical regime we have a complete timescale separation. 
Therefore each spreading of activity involves just clusters of 
connected critical sites. This is because the tree growth 
timescale $p^{-1}$ is much longer than the 
activity timescale, thus preventing that 
new critical sites change the connectivity properties of the 
configuration. 
In this conditions, the probability $P(s,\theta)$ 
to have an avalanche of size $s$ scales as the distribution
$n(s)$ of connected clusters with $s$ critical sites times 
the size of the cluster $s$. 
This factor takes into account the probability that the ignition 
process starts on any site of a cluster of size $s$. 
On the other hand,  the density of 
critical sites, leaving apart normalization factors, is given by
$\rho_c\sim \int sn(s) ds$ that is given  by the integral of the 
avalanche distribution. We can therefore write
\beq 
\Delta\rho_c\sim\int s^{-\tau}(1-{\cal G}(s/s_c(\theta)))ds
\eeq
where we used the explicit form of the avalanche distribution.
Noticing that ${\cal G}(s/s_c(\theta))\simeq 0$ for $s>s_c$ we 
obtain that the main contribution to the above integral is given by
\beq
\Delta\rho_c\sim\int_{s_c}^{\infty} s^{-\tau}ds,
\eeq
or as a result of the integration 
\beq
\Delta\rho_c\sim s_c^{1-\tau}.
\eeq
By substituting $s_c\sim\theta^{-1/\sigma}$ in the above expression and 
requiring the scaling consistency with Eq.(\ref{ffapp}), we finally 
obtain the scaling relation 
\beq
\zeta=\frac{\tau-1}{\sigma}.
\eeq

\newpage

\end{multicols}


\begin{thebibliography}{99}
\bibitem{nota1}
$J_{in}=h L^d$ is the average energy added considering just first order 
contribution in $h$ and $\rho_a$. In fact energy is not usually added 
on top of already active sites. This makes the energy addiction being
$J_{in}=h(1-\rho_a)L^d$. Also the energy balance in the stationary 
state is changed accordingly, thus obtaining 
$h(1-\rho_a)L^d=\epsilon\rho_aL^d$. The latter immediately gives the
stationary density of active sites as $\rho_a=h/(h+\epsilon)$.
For $\epsilon >>h$ we recover the first order result presented in the
text.  However higher order corrections can be considered, obtaining 
consistency relations useful for higher order treatment of the MF 
equations.
\bibitem{conn}
	The general form of the pair connectdness is
	$r^{2-d-\eta_{occ}}~\Gamma(r/\xi)$. In the MF description, where no 
	anomalous dimension appear, $\eta_{occ}=0$.
\bibitem{steps}
	It is easy to recognize that in dissipative model, each active 
	particle can wander only  for a finite number of steps.
	Each toppling process loses on average an amount $\epsilon$ of energy. 
	The energy unit needed to sustain the activity (wandering of active 
	site) is therefore dissipated after $\epsilon^{-1}$ toppling events
	(wandering steps). This indeed recover the result
	obtained for the average avalanche size in our MF approach.
\end{thebibliography}
\end{document}